\documentclass[3p,times,twocolumn]{elsarticle}
 \biboptions{comma,sort&compress}
 
\usepackage{graphicx}
\usepackage{here}
\usepackage{ecrc}

\volume{00}

\firstpage{1}

\journalname{Nuclear and Particle Physics Proceedings}

\runauth{}

\jid{nppp}

\jnltitlelogo{Nuclear and Particle Physics Proceedings}

\usepackage{amssymb}

\usepackage[figuresright]{rotating}

\def\beq{\begin{equation}}
\def\eeq{\end{equation}}
\def\bea{\begin{eqnarray}}
\def\eea{\end{eqnarray}}
\def\bq{\begin{quote}}
\def\eq{\end{quote}}

\def\nnb{\nonumber}
\def\ga{\left(}
\def\dr{\right)}

\def\nnb{\nonumber}
\def\la{\langle}
\def\ra{\rangle}
\def\nin{\noindent}
\def\ba{\vspace*{-0.2cm}\begin{array}}
\def\ea{\end{array}\vspace*{-0.2cm}}

\def\b{$\bullet~$}
\def\als{\alpha_s}

\def\gg2{ \la\alpha_s G^2 \ra}
\def\gg3{g^3f_{abc}\la G^aG^bG^c \ra}
\def\ggg4{\la\als^2G^4\ra}

\def\beq{\begin{equation}}
\def\enq{\end{equation}}
\def\beqa{\begin{eqnarray}}
\def\enqa{\end{eqnarray}}
\def\nnb{\nonumber}

\def\qq{\lag\bar{q}q\rag}

\def\lb{\label}

\newcommand{\rag}{\rangle}
\newcommand{\lag}{\langle}

\def\ln{\mbox{Log}}
\def\gg{\lag g^{2}_{s} G^2 \rag}
\def\ggg{\lag g^{3}_{s}G^3\rag}

\begin{document}

\begin{frontmatter}

\title{ Heavy-Light Exotics from QCD Laplace Sum Rules at N2LO in the chiral limit$^*$}
 \cortext[cor1]
 {Talk given at 19th International Conference in Quantum Chromodynamics (QCD 16,  4-8 july 2016, Montpellier - FR) and at the 8th International Conference in High-Energy Physics  (HEPMAD16-15th anniversary, 13-18th october 2016, Antananarivo, MG)}
 \author[label1]{R.M  Albuquerque}
\ead{ raphael.albuquerque@uerj.br}
\address[label1]{Faculty of Technology, Rio de Janeiro State University (FAT,UERJ), Brazil}
 \author[label2]{F. Fanomezana  \fnref{fn0} }
  \fntext[fn0] {PhD student.}

\ead{fanfenos@yahoo.fr}
\address[label2]{Institute of High-Energy Physics of Madagascar (iHEPMAD), University of Antananarivo, 
Madagascar}
\author[label3]{S. Narison}
    \ead{snarison@yahoo.fr}
    \address[label3]{Laboratoire
Particules et Univers de Montpellier, CNRS-IN2P3, 
Case 070, Place Eug\`ene
Bataillon, 34095 - Montpellier, France.}

 \author[label2]{A. Rabemananjara}
\ead{achris\_01@yahoo.fr}

 \author[label2]{D. Rabetiarivony\fnref{fn0}}
\ead{bidds.davidson@outlook.com} 

 \author[label2]{G.~Randriamanatrika\fnref{fn0}}
\ead{artesgaetan@gmail.com}

\pagestyle{myheadings}
\markright{ }
\begin{abstract}
These talks review and summarize our results in\,\cite{XYZ,X} on $XYZ$-like spectra obtained from QCD Laplace Sum Rules  in the chiral limit
at next-to-next-leading order (N2LO) of perturbation theory (PT) and including leading order (LO) contributions of dimensions 
$d\leq 6-8$ non-perturbative condensates. We conclude that the observed $XZ$ states are good candidates for $1^{+}$ and $0^+$ molecules or / and four-quark states while the predictions for $1^-$ and $0^-$ states are about 1.5 GeV above the $Y_{c,b}$ experimental candidates and hadronic thresholds. We (numerically) find that these exotic molecules couple weakly to the corresponding interpolating currents than ordinary $D,B$ heavy-light mesons while we observe that these couplings decrease faster [$1/m_b^{3/2}$ (resp. $1/m_b$) for the $1^+,0^+$ (resp. $1^-,0^-)$ states] than $1/m_b^{1/2}$. Our results do not also confirm the existence of the $X(5568)$ state
in agreement with LHCb findings. 
\end{abstract}

\end{frontmatter}
\vspace*{-1cm}
\section{Note for the Readers}
This paper summarizes the results in our original works\,\cite{XYZ,X}. 
Most of the references fairly quoted there are not repeated here due 
to space limitations. We sincerely apologize for that. 

 \vspace*{-0.5cm}
\section{Introduction}
 A large amount of  exotic hadrons which differ from the ``standard" $\bar cc$ charmonium and $\bar bb$ bottomium radial excitation states have been discovered in $D$ and $B$-factories through e.g. $J/\psi\pi^+\pi^-$ and $\Upsilon\pi^+\pi^-$ processes\,\cite{PDG,EXP}. They are referred as $XYZ$ states\,\cite{TH}.  In this talk, we shall present our predictions for the masses and couplings of these states obtained using the Laplace sum rule (LSR)\,\cite{SVZa,BELLa,BELLb,SNRAF} version of QCD spectral sum rules (QSSR)\,\cite{SVZa}\,\footnote{For a review, see e.g.\,\cite{SNB1,SNB2}.} known at next-to-next-leading order (N2LO) of PT series and including non-perturbative condensates of dimensions $d\leq 6-8$. In so doing, we assume a factorization of the four-quark spectral functions into a convolution of two ones built from quark bilinear currents as in \,\cite{PICH,BBAR2,BBAR3}. We show 
  in \,\cite{XYZ} that this factorization, though valid to leading order in $1/N_c$, can reproduce with a good accuracy the predictions for the masses and couplings  obtained from a complete lowest $\alpha_s$ order expression. 
 \section{Molecules and Four-quark two-point functions}
 
 We shall work with the transverse part $\Pi^{(1)}$ of the \\
 \newpage
 two-point spectral functions :
{   \footnotesize
 \bea
\Pi^{\mu\nu}(q)&\equiv&i\int d^4x ~e^{iq.x}\lag 0
|T[{\cal O}^\mu(x){\cal O}^{\nu\dagger}(0)]
|0\rag\nnb\\
&=&-\Pi^{(1)}(q^2)(g^{\mu\nu}-\frac{q^\mu q^\nu}{q^2})+\Pi^{(0)}(q^2)\frac{q^\mu
q^\nu}{ q^2}~,
\lb{2po}
\eea
}
for the spin 1 states while for the spin zero ones, we shall use 
 the  two-point functions $\psi^{(s,p)}(q^2)$ 
built directly from the (pseudo)scalar currents: 
{\footnotesize
\beq 
\psi^{(s,p)}(q^2)=i\int d^4x ~e^{iq.x}\lag 0
|T[{\cal O}^{(s,p)}(x){\cal O}^{(s,p)}(0)]
|0\rag~,
\label{2po5}
\eeq
}
which is related to $\Pi^{(0)}$ appearing in 
Eq.~(\ref{2po}) via  Ward identities\,\cite{SNB1,SNB2}.
\subsection*{\b Interpolating currents}
The interpolating currents ${\cal O}$ for the molecules (resp. four-quark states) are
given in Table\,\ref{tab:current} (resp.  Table\,\ref{tab:4qcurrent}).  
\vspace*{-0.5cm}
{\scriptsize
\begin{center}
\begin{table}[hbt]
\setlength{\tabcolsep}{0.5pc}
\newlength{\digitwidth} \settowidth{\digitwidth}{\rm 0}
\catcode`?=\active \def?{\kern\digitwidth}

 \caption{
Interpolating currents with a definite $C$-parity describing the molecule-like 
states. $Q\equiv$ $c$ (resp. $b$) for the $\bar DD$ (resp. $\bar BB$)-like molecules.  $q\equiv u,d$.  
}
    {\footnotesize
\begin{tabular}{lcl}

&\\
\hline
\hline
States& $J^{PC}$&Molecule Currents  $\equiv{\cal O}_{mol}(x)$  \\
\hline
&$\bf 0^{++}$& \\
$\bar DD,~\bar BB$  &
&$( \bar{q} \gamma_5 Q ) (\bar{Q} \gamma_5 q)$ \\
%
$\bar D^*D^*,\bar B^*B^*$ & 
&$( \bar{q} \gamma_\mu Q ) (\bar{Q} \gamma^\mu q)$ \\ 
$\bar D^*_0D^*_0,~\bar B_0^*B^*_0$&  & $( \bar{q} Q ) (\bar{Q} q)$\\ 
&$\bf 1^{++}$& \\
$\bar D^*D,~\bar B^*B$& 
&$ \frac{i}{\sqrt{2}} \Big[ (\bar{Q} \gamma_\mu q) ( \bar{q} \gamma_5 Q ) 
     - (\bar{q} \gamma_\mu Q) ( \bar{Q} \gamma_5 q ) \Big]$\\ 
 $\bar D^*_0D_1,~\bar B^*_0B_1$  & 
&$ \frac{1}{\sqrt{2}} \Big[ ( \bar{q} Q ) (\bar{Q} \gamma_\mu \gamma_5 q)
     + ( \bar{Q} q ) (\bar{q} \gamma_\mu \gamma_5 Q) \Big]$\\ 
&$\bf 0^{-\pm}$& \\
     $\bar D^*_0D,~\bar B_0^*B$ & & $ \frac{1}{\sqrt{2}} \Big[ ( \bar{q} Q ) (\bar{Q} \gamma_5 q) 
     \pm ( \bar{Q} q ) (\bar{q} \gamma_5 Q) \Big]$\\
$\bar D^*D_1,~\bar B^*B_1$ & 
&$ \frac{1}{\sqrt{2}} \Big[ ( \bar{Q} \gamma_\mu q ) (\bar{q} \gamma^\mu \gamma_5 Q)
     \mp ( \bar{Q} \gamma_\mu \gamma_5 q ) (\bar{q} \gamma^\mu Q) \Big]$\\
&$\bf 1^{-\pm}$& \\
$\bar D^*_0D^*,~\bar B^*_0B^*$  & 
&$ \frac{1}{\sqrt{2}} \Big[ ( \bar{q} Q ) (\bar{Q} \gamma_\mu q) 
     \mp ( \bar{Q} q ) (\bar{q} \gamma_\mu Q) \Big]$\\ 
     $\bar DD_1,~\bar BB_1$  &
 &    $ \frac{i}{\sqrt{2}} \Big[ ( \bar{Q} \gamma_\mu \gamma_5 q ) (\bar{q} \gamma_5 Q)
     \pm ( \bar{q} \gamma_\mu \gamma_5 Q ) (\bar{Q} \gamma_5 q) \Big]$\\
\hline
\hline
\end{tabular}
}
\label{tab:current}
\end{table}
\end{center}
}
\vspace*{-1.5cm}
{\scriptsize
\begin{center}
\begin{table}[hbt]
\setlength{\tabcolsep}{0.2pc}

 \caption{
Interpolating currents with a definite $P$-parity describing the four-quark 
states. $Q\equiv$ $c$ (resp. $b$) in the charm and bottom channels.  $q\equiv u,d$.  
}
    {\footnotesize
\begin{tabular}{ll}

&\\
\hline
\hline
$J^{P}$&Four-Quark Currents  $\equiv{\cal O}_{4q}(x)$  \\
\hline
$\bf 0^{+}$&$\epsilon_{abc}\epsilon_{dec}  \bigg[
		\big( q^T_a \: C\gamma_5 \:Q_b \big) \big( \bar{q}_d \: \gamma_5 C \: \bar{Q}^T_e \big) 
		+ k \big( q^T_a \: C \:Q_b \big) \big( \bar{q}_d \: C \: \bar{Q}^T_e \big) \bigg] $\\		
$\bf 1^{+}$& $\epsilon_{abc}\epsilon_{dec}  \bigg[
		\big( q^T_a \: C\gamma_5 \:Q_b \big) \big( \bar{q}_d \: \gamma_\mu C \: \bar{Q}^T_e \big) 
		+ k \big( q^T_a \: C \:Q_b \big) \big( \bar{q}_d \: \gamma_\mu\gamma_5 C \: \bar{Q}^T_e \big) \bigg]  $ \\
	
$\bf 0^{-}$&	$\epsilon_{abc}\epsilon_{dec}  \bigg[
		\big( q^T_a \: C\gamma_5 \:Q_b \big) \big( \bar{q}_d \: C \: \bar{Q}^T_e \big) 
		+ k \big( q^T_a \: C \:Q_b \big) \big( \bar{q}_d \:\gamma_5 C \: \bar{Q}^T_e \big) \bigg] $\\	
$\bf 1^{-}$&$  \epsilon_{abc}\epsilon_{dec}  \big[
		\big( q^T_a \: C\gamma_5 \:Q_b \big) \big( \bar{q}_d \: \gamma_\mu\gamma_5 C \: \bar{Q}^T_e \big) 
		+ k \big( q^T_a \: C \:Q_b \big) \big( \bar{q}_d \: \gamma_\mu C \: \bar{Q}^T_e \big) \big]$\\
\hline
\hline
\end{tabular}
}
\label{tab:4qcurrent}
\end{table}
\end{center}
}
\nin
\vspace*{-0.5cm}
\nin
\subsection*{\b Spectral Function within MDA}

 We shall  use the Minimal Duality Ansatz (MDA) given in Eq. \ref{eq:duality} for parametrizing the spectral function:
 {\footnotesize
\beq
\frac{1}{\pi}\mbox{ Im}\Pi(t)\simeq f^2_{H}M_H^8\delta(t-M_H^2)
  \ + \
  ``\mbox{QCD continuum}" \theta (t-t_c),
\label{eq:duality}
\eeq
}
where $f_H$ is the decay constant defined as:
{\footnotesize
\beq
\la 0| {\cal O}^{(s,p)}|H\ra=f^{(s,p)_H}M^4_{H}~,~~~~~~~~~~~~\la 0| {\cal O}^\mu |H\ra=f^{(1)}_{H}M^5_{H}\epsilon_\mu~,
\label{eq:coupling}
\eeq
}
respectively for spin 0 and 1 hadronic states $H$ with  $\epsilon_\mu$ the vector polarization.
The higher states contributions are smeared by the ``QCD continuum" coming from the discontinuity of the QCD diagrams and starting from a constant threshold $t_c$.  
\subsection*{\b NLO and N2LO PT corrections using factorization}
Assuming a factorization of the four-quark interpolating current as a natural consequence of the molecule
definition of the state, we can write the corresponding spectral function as a convolution of the
spectral functions associated to quark bilinear current
for the  $\bar DD^*$ and $\bar D^*_0D^*$ spin 1 states:
{\footnotesize
\bea
\frac{1}{ \pi}{\rm Im} \Pi^{(1)}_{H}(t)&=& \theta (t-4M_Q^2)\ga \frac{1}{ 4\pi}\dr^2 t^2 \int_{M_Q^2}^{(\sqrt{t}-M_Q)^2}\hspace*{-0.5cm}dt_1\int_{M_Q^2}^{(\sqrt{t}-\sqrt{t_1})^2} \hspace*{-1cm}dt_2\nnb\\
&&\times~\lambda^{3/2}\frac{1}{ \pi}{\rm Im} \Pi^{(1)}(t_1) \frac{1}{ \pi}{\rm Im} \psi^{(s,p)}(t_2)~.
\label{eq:convolution}
\eea
}
For the  $\bar DD$ spin 0 state, one has:
{\footnotesize
\bea
\frac{1}{ \pi}{\rm Im} \psi^{(s)}_{H}(t)&=& \theta (t-4M_Q^2)\ga \frac{1}{4\pi}\dr^2 t^2 \int_{m_Q^2}^{(\sqrt{t}-M_Q)^2}\hspace*{-0.5cm}dt_1\int_{m_Q^2}^{(\sqrt{t}-\sqrt{t_1})^2}  \hspace*{-1cm}dt_2~\nnb\\
&&\times~\lambda^{1/2}\ga \frac{t_1}{ t}+ \frac{t_2}{ t}-1\dr^2\nnb\\
&&\times ~\frac{1}{ \pi}{\rm Im}\psi^{(p)}(t_1) \frac{1}{ \pi} {\rm Im} \psi^{(p)}(t_2),
\eea
}
and for the $\bar D^*D^*$ spin 0 state:
{\footnotesize
\bea
\frac{1}{ \pi}{\rm Im} \psi_{H}(t)&=& \theta (t-4M_Q^2)\ga \frac{1}{4\pi}\dr^2 t^2 \int_{m_Q^2}^{(\sqrt{t}-M_Q)^2}\hspace*{-0.5cm}dt_1\int_{m_Q^2}^{(\sqrt{t}-\sqrt{t_1})^2}  \hspace*{-1cm}dt_2~\nnb\\
&&\times~\lambda^{1/2}\Big{[}\ga \frac{t_1}{ t}+ \frac{t_2}{ t}-1\dr^2
+\frac{8t_1t_2}{ t^2}\Big{]}\nnb\\
&&\times ~\frac{1}{ \pi}{\rm Im} \Pi^{(1)}(t_1) \frac{1}{ \pi} {\rm Im} \Pi^{(1)}(t_2),
\eea
}
where:
{\footnotesize
\beq
\lambda=\ga 1-\frac{\ga \sqrt{t_1}- \sqrt{t_2}\dr^2}{ t}\dr \ga 1-\frac{\ga \sqrt{t_1}+ \sqrt{t_2}\dr^2}{ t}\dr~,
\eeq
}
is the phase space factor and $M_Q$ is the on-shell heavy quark mass. 
Im $ \Pi^{(1)}(t)$ is the spectral function associated to the bilinear $\bar c\gamma_\mu (\gamma_5)q$  vector or axial-vector current, while Im $\psi^{(5)}(t)$ is associated to the 
$\bar c(\gamma_5)q$  scalar or pseudoscalar current\,\footnote{In the chiral limit $m_q=0$, the PT expressions of the vector (resp. scalar) and axial-vector (resp. pseudoscalar) spectral functions are the same.}. An analogous convolution is assumed for the four-quark states.
\subsection*{\b The Laplace sum rule (LSR)}
The exponential or Laplace sum rule (LSR) and its ratio read\,\footnote{The last equality in Eq.\,\ref{eq:ratioLSR} is obtained when one uses MDA in Eq.\,\ref{eq:duality} for parametrizing the spectral function.}:
{\footnotesize
\beq
{\cal L}_{H}(\tau,t_c,\mu)=\int_{4M_Q^2}^{t_c}dt~e^{-t\tau}\frac{1}{\pi} \mbox{Im}\Pi^{(1,0)}_{H}(t,\mu)~,
\label{eq:LSR}
\eeq
\beq\label{eq:ratioLSR}
{\cal R}_{H}(\tau,t_c,\mu) = \frac{\int_{4M_Q^2}^{t_c} dt~t~ e^{-t\tau}\frac{1}{\pi}\mbox{Im}\Pi^{(1,0)}_{H}(t,\mu)}
{\int_{4M_Q^2}^{t_c} dt~ e^{-t\tau} \frac{1}{\pi} \mbox{Im}\Pi^{(1,0)}_{H}(t,\mu)}\simeq M_R^2~,
\eeq
}
where $\mu$ is the subtraction point which appears in the approximate QCD series when radiative corrections are included and $\tau$ is the sum rule variable replacing $q^2$. 
Similar sum rules are obtained for the (pseudo)scalar two-point function $\psi^{(s,p)}(q^2)$. 
\subsection*{\b Stability criteria and some phenomenological tests}
The variables $\tau,\mu$ and $t_c$ are, in principle, free parameters. 
We shall use stability criteria (if any), with respect to these free 3 parameters,  for extracting the optimal results. 
In the standard Minimal Duality Ansatz (MDA) given in Eq. \ref{eq:duality} for parametrizing the spectral function,
 the ``QCD continuum" threshold $t_c$ is constant and is independent on the subtraction point $\mu$. One should notice that this standard MDA with constant $t_c$ describes quite well the properties of the lowest ground state as explicitly demonstrated in \cite{SNFB12a} and in various examples\,\cite{SNB1,SNB2} after confronting the  integrated spectral function within this simple parametrization with the full data measurements. It has been also successfully tested in the large $N_c$ limit of QCD in \cite{PERISb}. 
Though it is difficult to estimate with a good precision the systematic error related to this simple model, these features indicate the ability of the model for reproducing accurately the data. We expect that the same feature is reproduced for the case of the XYZ discussed here where complete data are still lacking.
\section{QCD input parameters}
The QCD parameters which shall appear in the following analysis will be the charm and bottom quark masses $ m_{c,b}$ (we shall neglect  the light quark masses $q\equiv u,d$),
the light quark condensate $\qq$,  the gluon condensates $ \lag
\alpha_sG^2\rag
\equiv \la \alpha_s G^a_{\mu\nu}G_a^{\mu\nu}\ra$ 
and $ \la g^3G^3\ra
\equiv \la g^3f_{abc}G^a_{\mu\nu}G^{b,\nu}_{\rho}G^{c,\rho\mu}\ra$, 
the mixed condensate $\la\bar qGq\ra
\equiv {\la\bar qg\sigma^{\mu\nu} (\lambda_a/2) G^a_{\mu\nu}q\ra}=M_0^2\la \bar qq\ra$ 
and the four-quark 
 condensate $\rho\alpha_s\la\bar qq\ra^2$, where
 $\rho\simeq 3-4$ indicates the deviation from the four-quark vacuum 
saturation. Their values are given in Table \ref{tab:param}. We shall work with the running
light quark condensates and masses, which read to leading order in $\alpha_s$: 
{\footnotesize
\beq
{\la\bar qq\ra}(\tau)=-{\hat \mu_q^3  \ga-\beta_1a_s\dr^{2/{
\beta_1}}},
{\la\bar q Gq\ra}(\tau)=-{M_0^2{\hat \mu_q^3} \ga-\beta_1a_s\dr^{1/{3\beta_1}}},
\label{d4g}
\eeq
}
where $\beta_1=-(1/2)(11-2n_f/3)$ is the first coefficient of the $\beta$ function 
for $n_f$ flavours; $a_s\equiv \alpha_s(\tau)/\pi$; 
$\hat\mu_q$ is the spontaneous RGI light quark condensate \cite{FNR}. 
{\scriptsize
\begin{table}[hbt]
\setlength{\tabcolsep}{0.45pc}
 \caption{QCD input parameters:
the original errors for 
$\la\alpha_s G^2\ra$, $\la g^3  G^3\ra$ and $\rho \la \bar qq\ra^2$ have been multiplied by about a factor 3 for a conservative estimate of the errors (see also the text).}  
    {\footnotesize
  {\begin{tabular}{lll}
&\\
\hline
\hline
Parameters&Values& Ref.    \\
\hline
$\alpha_s(M_\tau)$& $0.325(8)$&\cite{BNPa,SNTAU,PICHTAU}\\
$\overline{m}_c(m_c)$&$1261(12)$ MeV &average \cite{SNH10a,PDG,IOFFEa}\\
$\overline{m}_b(m_b)$&$4177(11)$ MeV&average \cite{SNH10a,PDG}\\
$\hat \mu_q$&$(253\pm 6)$ MeV&\cite{SNB1,SNB2,SNmassa,SNmassb,SNmass98a}\\
$M_0^2$&$(0.8 \pm 0.2)$ GeV$^2$&\cite{JAMI2a,JAMI2c,HEIDa,HEIDc,SNhl}\\
$\la\alpha_s G^2\ra$& $(7\pm 3)\times 10^{-2}$ GeV$^4$&
\cite{SNTAU,LNT,SNIa,YNDU,BELLa,BELLb,SNH10a,SNG1}\\
$\la g^3  G^3\ra$& $(8.2\pm 2.0)$ GeV$^2\times\la\alpha_s G^2\ra$&
\cite{SNH10a}\\
$\rho \alpha_s\la \bar qq\ra^2$&$(5.8\pm 1.8)\times 10^{-4}$ GeV$^6$&\cite{SNTAU,LNT,SNT,JAMI2a,JAMI2c}\\
\hline\hline
\end{tabular}}
}
\label{tab:param}
\end{table}
} 
\section{QCD expressions of the spectral functions}
In our works\,\cite{XYZ,X}, we provide new compact integrated expressions of the
spectral functions at LO of PT QCD and including non-perturbative condensates
having dimensions $d\leq 6-8$. NLO and N2LO corrections are introduced using 
the convolution integrals in Eq.\,\ref{eq:convolution}. The expressions of spectral functions of heavy-light bilinear currents are known to order $\alpha_s$ (NLO) from \cite{BROAD} and to order $\alpha_s^2$ (N2LO) from \cite{CHETa} which are available as a Mathematica Program named  Rvs. N3LO corrections are estimated  from the geometric growth of the QCD PT series \cite{SNZ} as a source of the PT errors, which we expect to give a good approximation of the uncalculated higher order terms dual to the $1/q^2$ contribution of a tachyonic gluon mass\,\cite{CNZ1,CNZ2} (for reviews see e.g\,\cite{ZAK1,ZAK2}).  

In our analysis, we replace the on-shell (pole) mass appearing in the LO spectral functions with the running mass using the relation, to order $\alpha_s^2$ \cite{TAR,COQUEa,COQUEb,SNPOLEa,SNPOLEb,BROAD2a,AVDEEV,BROAD2b,CHET2a,CHET2b}:
{\footnotesize
\bea
M_Q &=& \overline{m}_Q(\mu)\Big{[}
1+\frac{4}{3} a_s+ (16.2163 -1.0414 n_l)a_s^2\nnb\\
&&+\ln{\ga\frac{\mu}{ M_Q}\dr^2} \ga a_s+(8.8472 -0.3611 n_l) a_s^2\dr\nnb\\
&&+\ln^2{\ga\frac{\mu}{ M_Q}\dr^2} \ga 1.7917 -0.0833 n_l\dr a_s^2...\Big{]},
\label{eq:pole}
\eea
}
for $n_l$ light flavours where $\mu$ is the arbitrary subtraction point and $a_s\equiv \alpha_s/\pi$.
\section{Tests of the Factorization Assumption}\label{sec:factor}
\subsection* {\b $\bar D^*_0D^*(1^-)$  molecule state at LO}
\begin{figure}[hbt] 
\begin{center}
{\includegraphics[width=6.29cm  ]{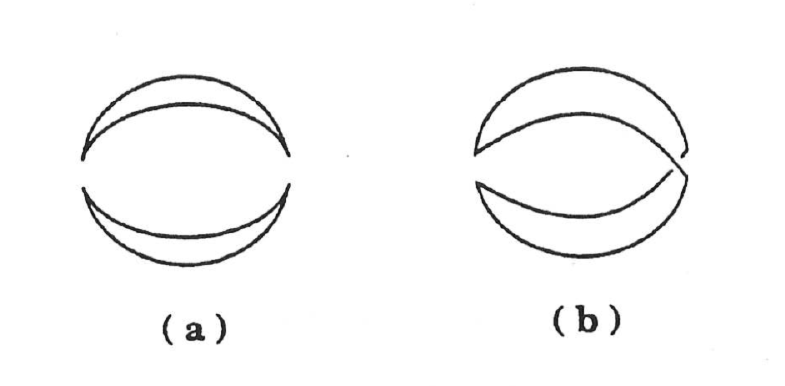}}
\caption{
\scriptsize 
{\bf (a)} Factorized contribution to the four-quark correlator at lowest order of PT; {\bf (b)} Non-factorized contribution at lowest order of PT (the figure comes from\,\cite{PICH}).
}
\label{fig:factor} 
\end{center}
\end{figure}
\begin{figure}[hbt] 
\begin{center}
{\includegraphics[width=3.85cm  ]{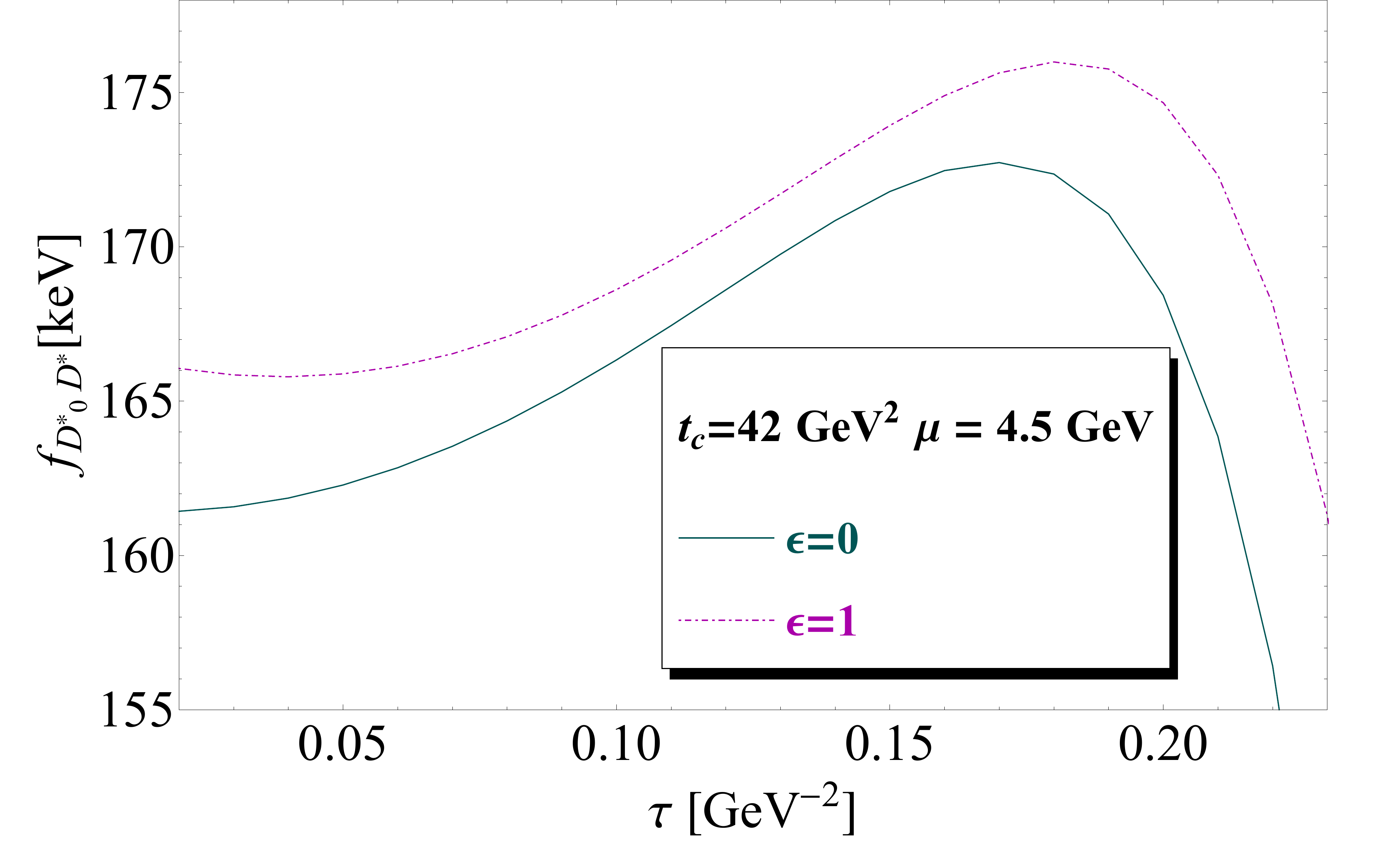}}
{\includegraphics[width=3.85cm  ]{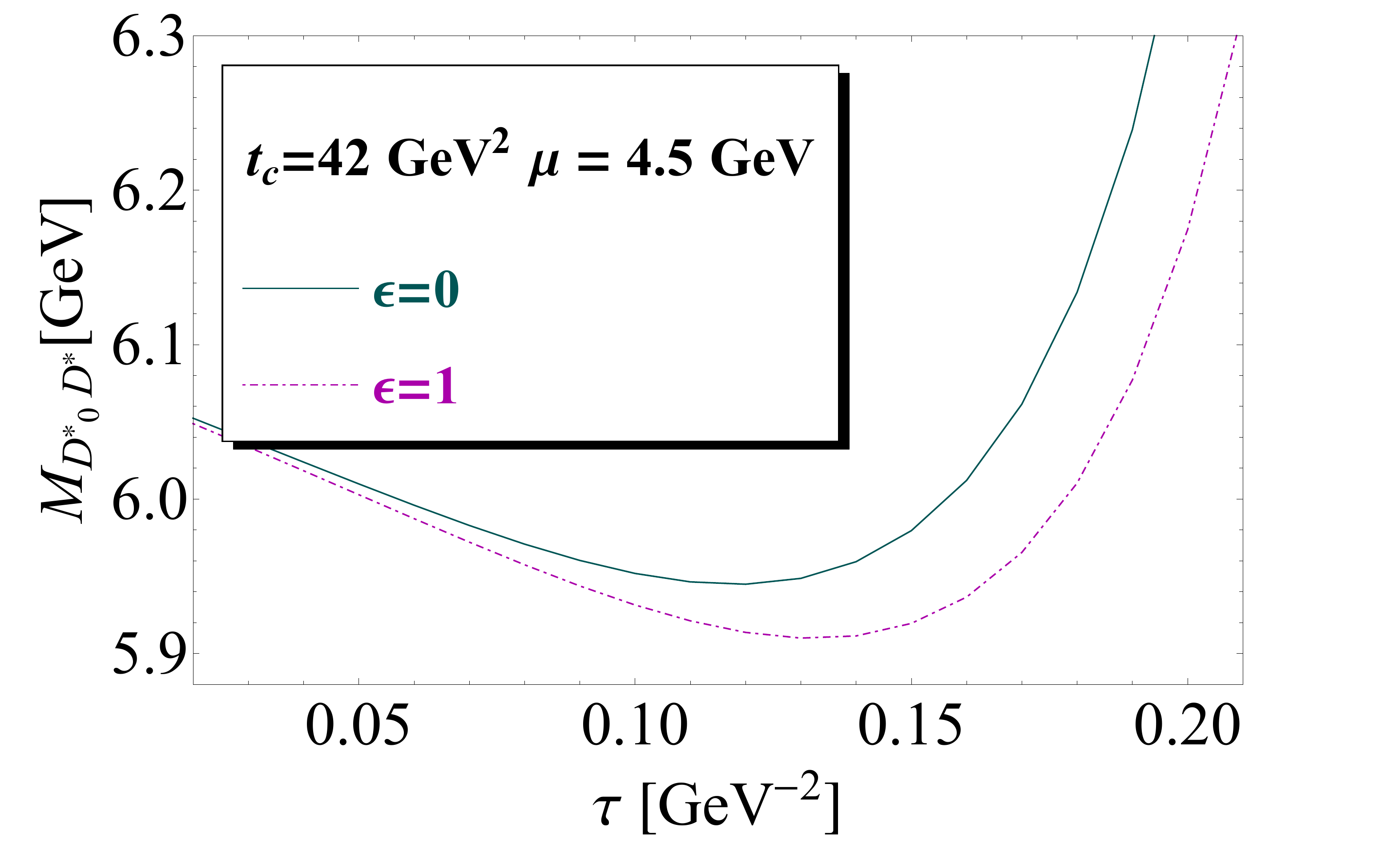}}
\scriptsize\centerline {\hspace*{-1cm} a)\hspace*{4cm} b) 
}
\caption{
\scriptsize 
{\bf a)} Factorized ($\epsilon=0$) and full ($\epsilon=1$) lowest order PT$\oplus$NP contributions to $f_{D^*_0D^*}$  as function of $\tau$ for a given value of $t_c=42$ GeV$^2$, $\mu=4.5$ GeV, $\overline{m}_c(\overline{m}_c)=1.26$ GeV and using the QCD parameters in Table\,\ref{tab:param}; {\bf b)} The same as a) but for the mass $M_{D^*_0D^*}$.
}
\label{fig:dstar0dstar-b01a} 
\end{center}
\end{figure}

 
\nin
 In the following, we shall test the factorization assumption if one does it at lowest order (LO) of perturbation theory (PT) by taking the example of the  $\bar D^*_0D^*(1^-)$  molecule state. To LO of PT, the four-quark correlator can be subdivided  into its factorized (Fig.\,\ref{fig:factor}a) and its non-factorized (Fig.\,\ref{fig:factor}b) parts.The analysis for the decay constant and mass including NP contributions up to dimension $d=6$ is shown in Fig.\,{\ref{fig:dstar0dstar-b01a}}. 
\nin
We conclude from the previous two examples that assuming a factorization of the PT at LO and including NP contributions induces an effect  about 2.2\% for the 
decay constant and 0.5\% for the mass which is quite tiny. However, to avoid this (small) effect, we shall work in the following with the full non-factorized PT$\oplus$NP of the  LO expressions. 
\subsection*{\b $B^0\bar B^0$ four-quark correlator at NLO}
\begin{figure}[hbt] 
\begin{center}
{\includegraphics[width=6.29cm  ]{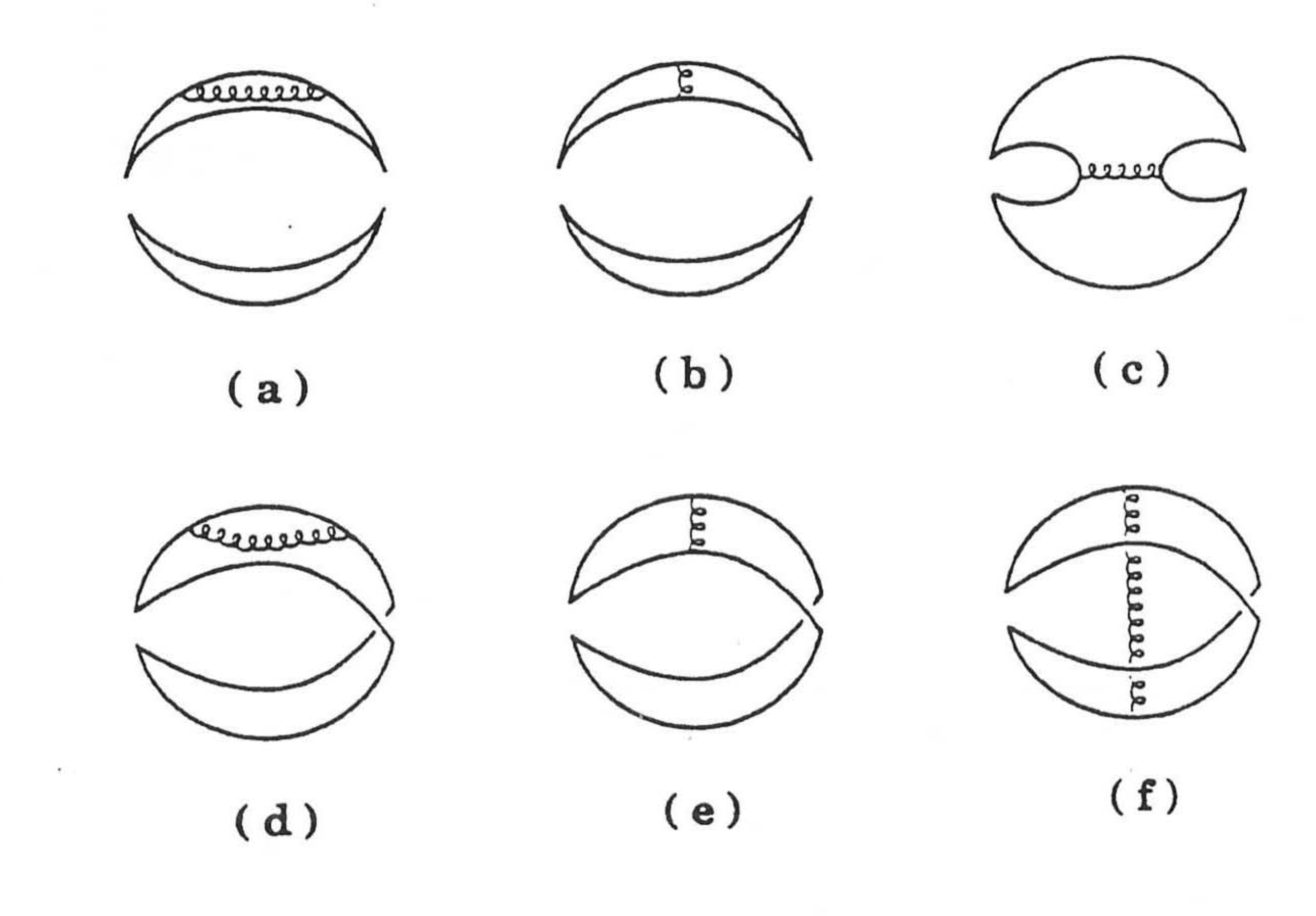}}
\caption{
\scriptsize 
{\bf (a,b)} Factorized contributions to the four-quark correlator at NLO of PT; {\bf (c to f)} Non-factorized contributions  at NLO of PT (the figure comes from\,\cite{PICH}).
}
\label{fig:factoras} 
\end{center}
\end{figure} 
\nin
For extracting the PT $\alpha_s^n$ corrections to the correlator and due to the technical complexity of the calculations, we shall assume that these radiative corrections are dominated by the ones from the factorized diagrams
(Fig.\,\ref{fig:factoras}a,b) 
while we neglect the ones from non-factorized ones 
(Fig.\,\ref{fig:factoras}c to f). 
This fact has been proven explicitly by \,\cite{BBAR2,BBAR3}  in the case of the $\bar B^0B^0$ systems (very similar correlator as the ones discussed in the following) where the non-factorized $\alpha_s$ corrections do not exceed 10\% of the total $\alpha_s$ contributions. 
\subsection*{\b Conclusions} 
\nin
We expect from the previous LO example that the masses of the molecules are known with a good accuracy while, for the coupling, we shall have in mind the systematics induced by the radiative corrections estimated by keeping only the factorized diagrams. The contributions of the factorized diagrams will be extracted from the convolution integrals given in Eq.\,\ref{eq:convolution}. Here, due to a partial cancellation of the corrections, the suppression of the NLO corrections will be more pronounced in the extraction of the meson masses from the ratio of sum rules than to the case of the $\bar B^0B^0$ systems. 
\section{$ \bar DD$ molecule decay constant and mass}

\subsection*{\b $\tau$ and $t_c$ stabilities}
\begin{figure}[hbt] 
\begin{center}
{\includegraphics[width=3.85cm  ]{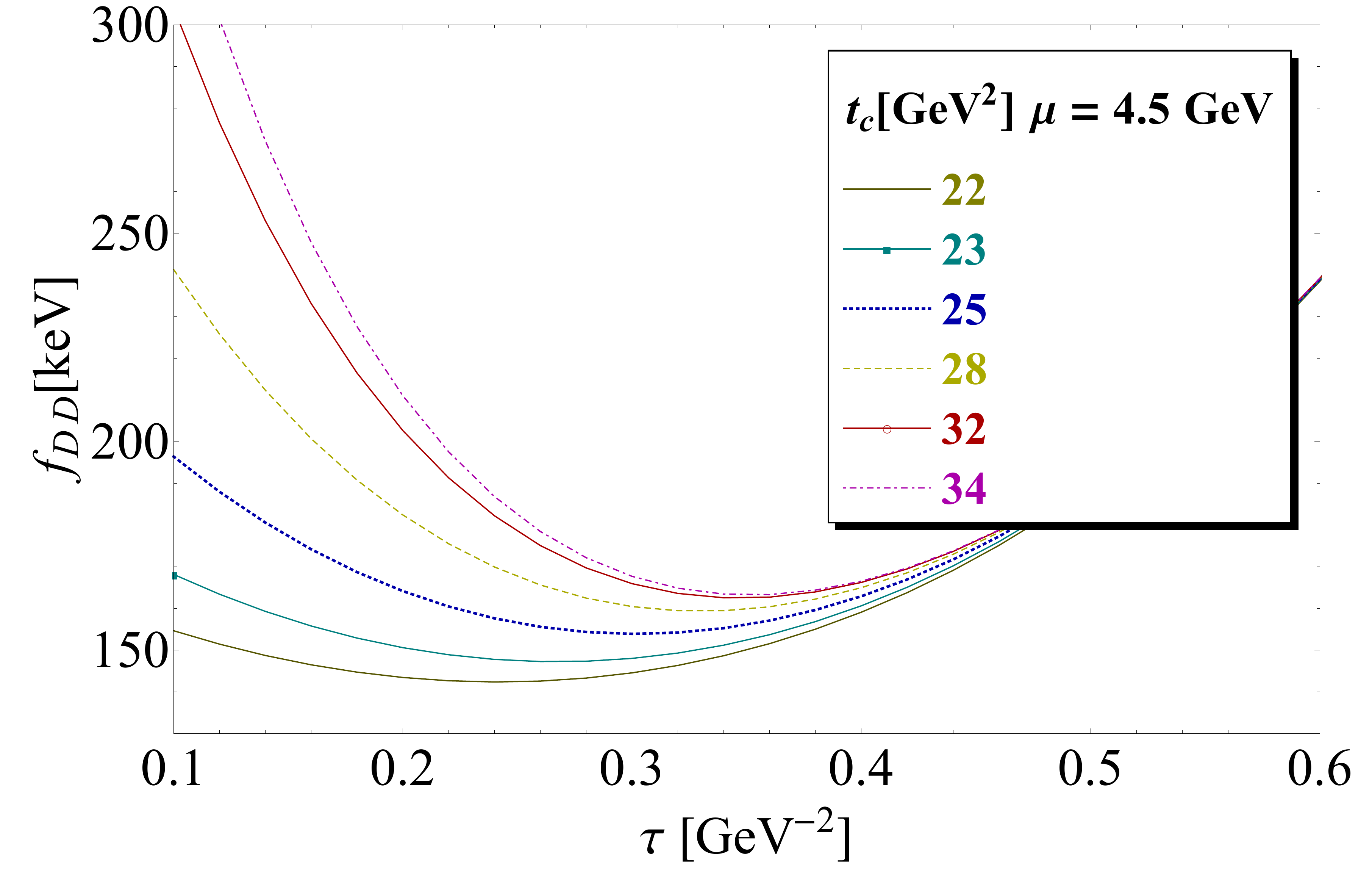}}
{\includegraphics[width=3.85cm  ]{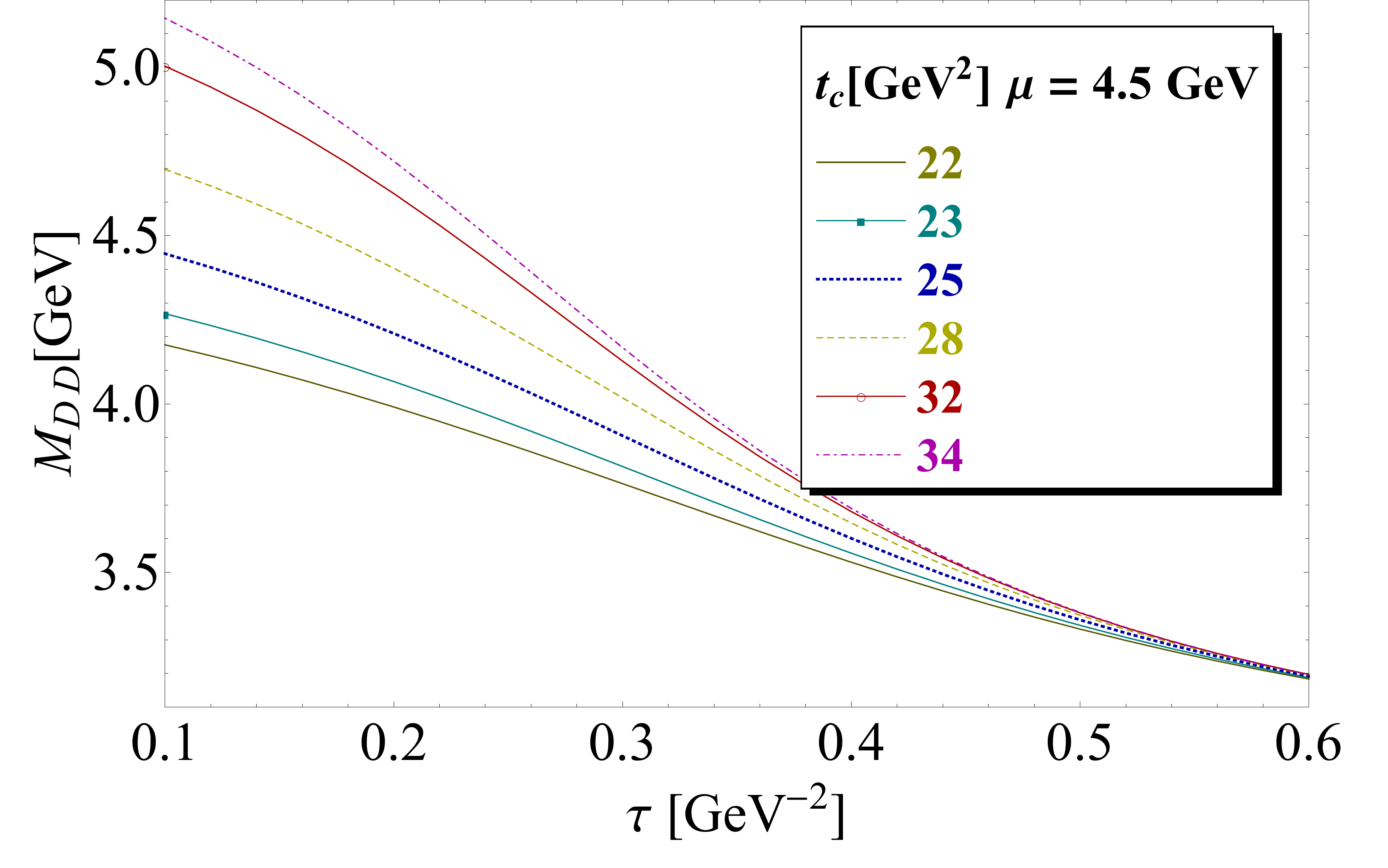}}
\scriptsize\centerline {\hspace*{-1cm} a)\hspace*{4cm} b) }
\caption{
\scriptsize 
{\bf a)} $f_{DD}$  at LO as function of $\tau$ for different values of $t_c$, for $\mu=4.5$ GeV  and for the QCD parameters in Table\,\ref{tab:param}; {\bf b)} The same as a) but for the mass $M_{DD}$.
}
\label{fig:d-lo} 
\end{center}
\end{figure}

\begin{figure}[hbt] 
\begin{center}
{\includegraphics[width=3.85cm  ]{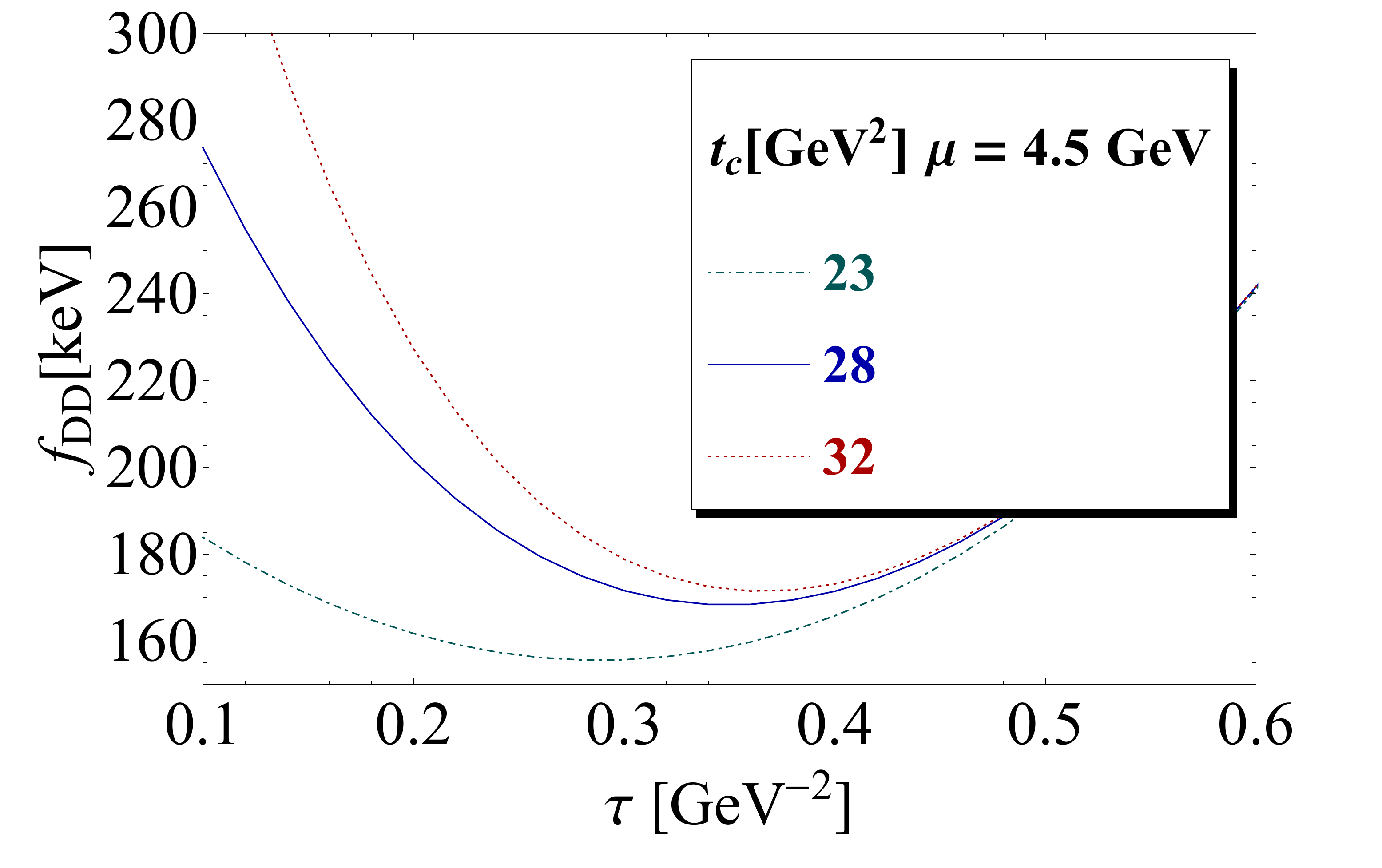}}
{\includegraphics[width=3.85cm  ]{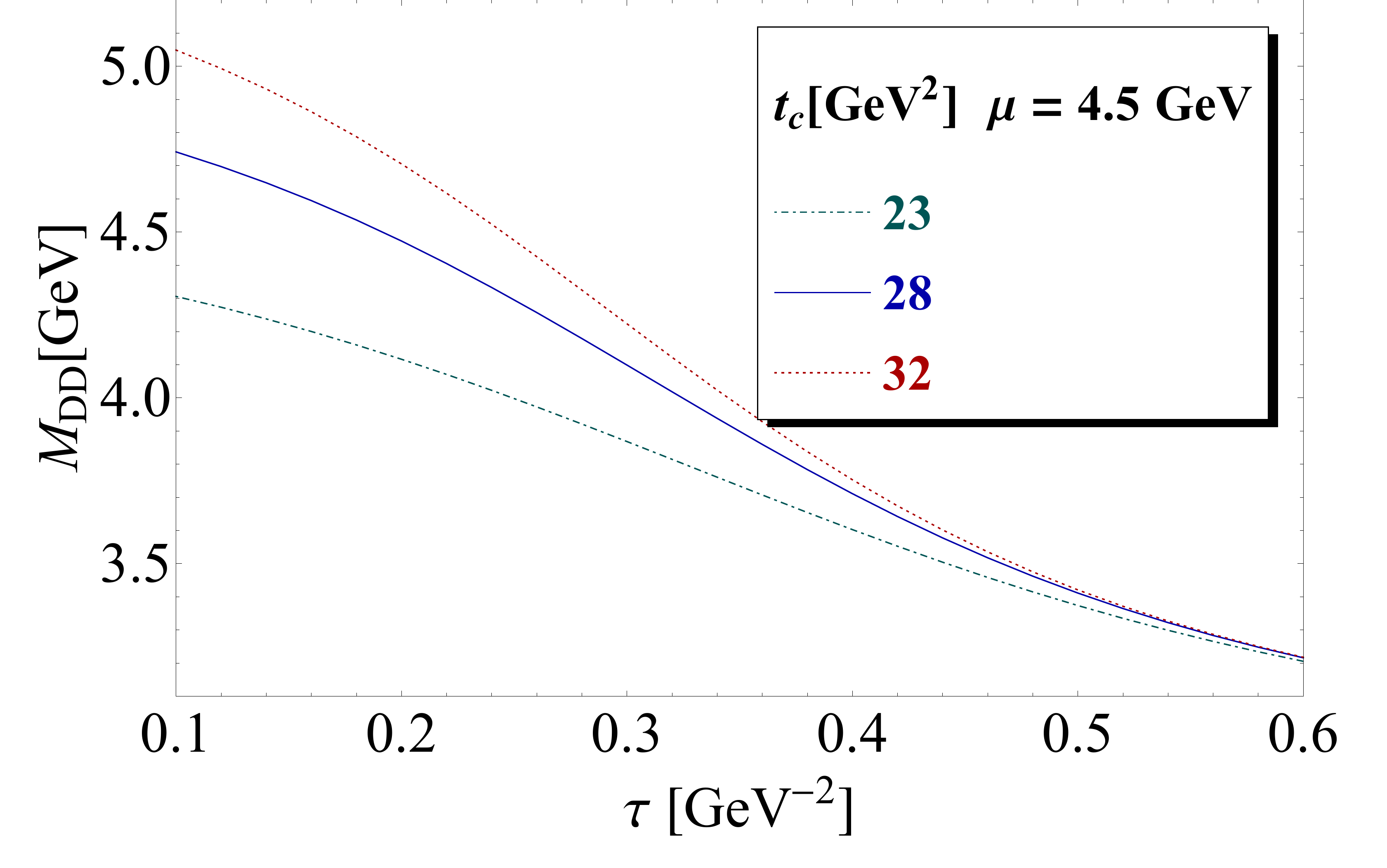}}
\scriptsize\centerline {\hspace*{-1cm} a)\hspace*{4cm} b)  }
\caption{
\scriptsize 
{\bf a)} $f_{DD}$ at NLO  as function of $\tau$ for different values of $t_c$, for $\mu=4.5$ GeV  and for the QCD parameters in Table\,\ref{tab:param}; {\bf b)} The same as a) but for the mass $M_{DD}$.
}
\label{fig:d-nlo} 
\end{center}
\end{figure}

\begin{figure}[hbt] 
\begin{center}
{\includegraphics[width=3.85cm  ]{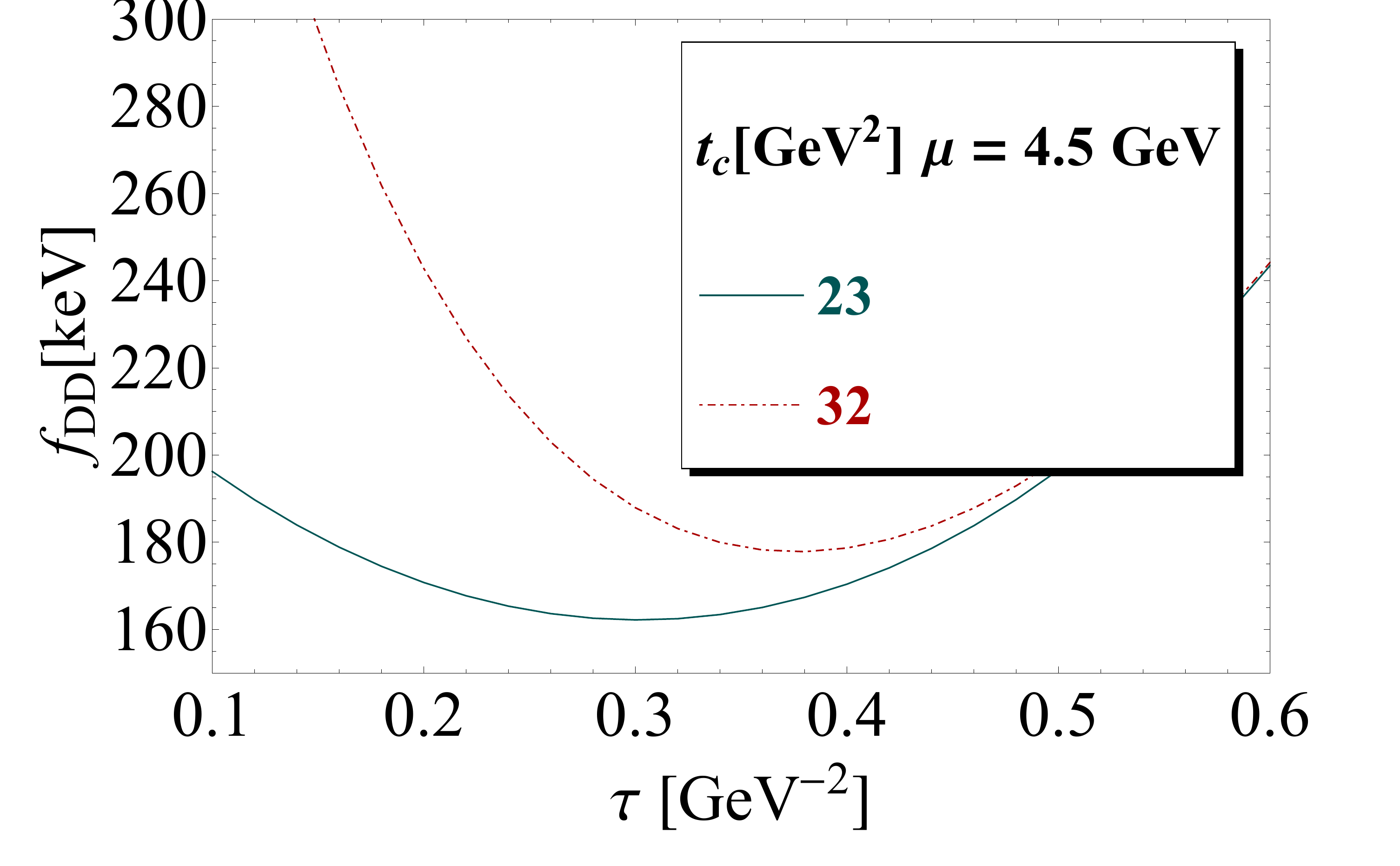}}
{\includegraphics[width=3.85cm  ]{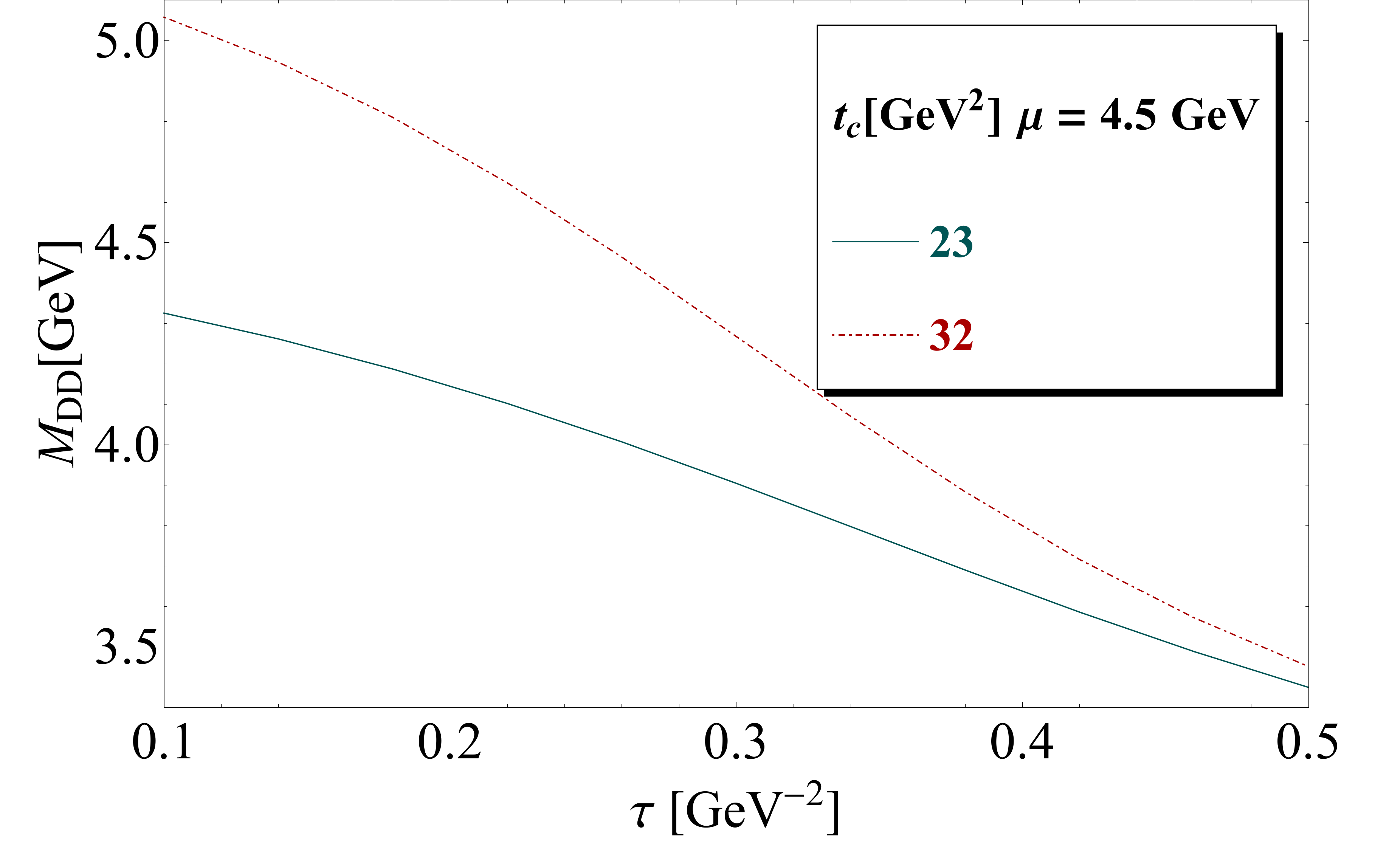}}
\scriptsize\centerline {\hspace*{-1cm} a)\hspace*{4cm} b)  }
\caption{
\scriptsize 
{\bf a)} $f_{DD}$ at N2LO  as function of $\tau$ for different values of $t_c$, for $\mu=4.5$ GeV  and for the QCD parameters in Table\,\ref{tab:param}; {\bf b)} The same as a) but for the mass $M_{DD}$.
}
\label{fig:d-n2lo} 
\end{center}
\end{figure}

\nin
 We study the behavior of the coupling\,\footnote{Here and in the following : decay constant is the same as : coupling.} $f_{DD}$  and mass $M_{DD}$ in terms of LSR variable $\tau$ at different values of $t_c$ as shown in Fig.\ref{fig:d-lo} at LO, in Fig.\,\ref{fig:d-nlo} at NLO  and in Fig.\,\ref{fig:d-n2lo} at N2LO.
\nin
 We consider, as a final and conservative result, the one corresponding to the beginning of the $\tau$-stability ($\tau\simeq$ 0.25 GeV$^{-2}$) for $t_c$=22 GeV$^2$ until the one where $t_c$-stability starts to be reached for $t_c\simeq$ 32 GeV$^2$ and for $\tau\simeq$ 0.35 GeV$^{-2}$. In these stability regions, the requirement that the pole contribution is larger than the one of the continuum  is automatically satisfied.
\nin
\nin
   \subsection*{\b Running versus the pole quark mass definitions}
\begin{figure}[hbt] 
\begin{center}
{\includegraphics[width=3.85cm  ]{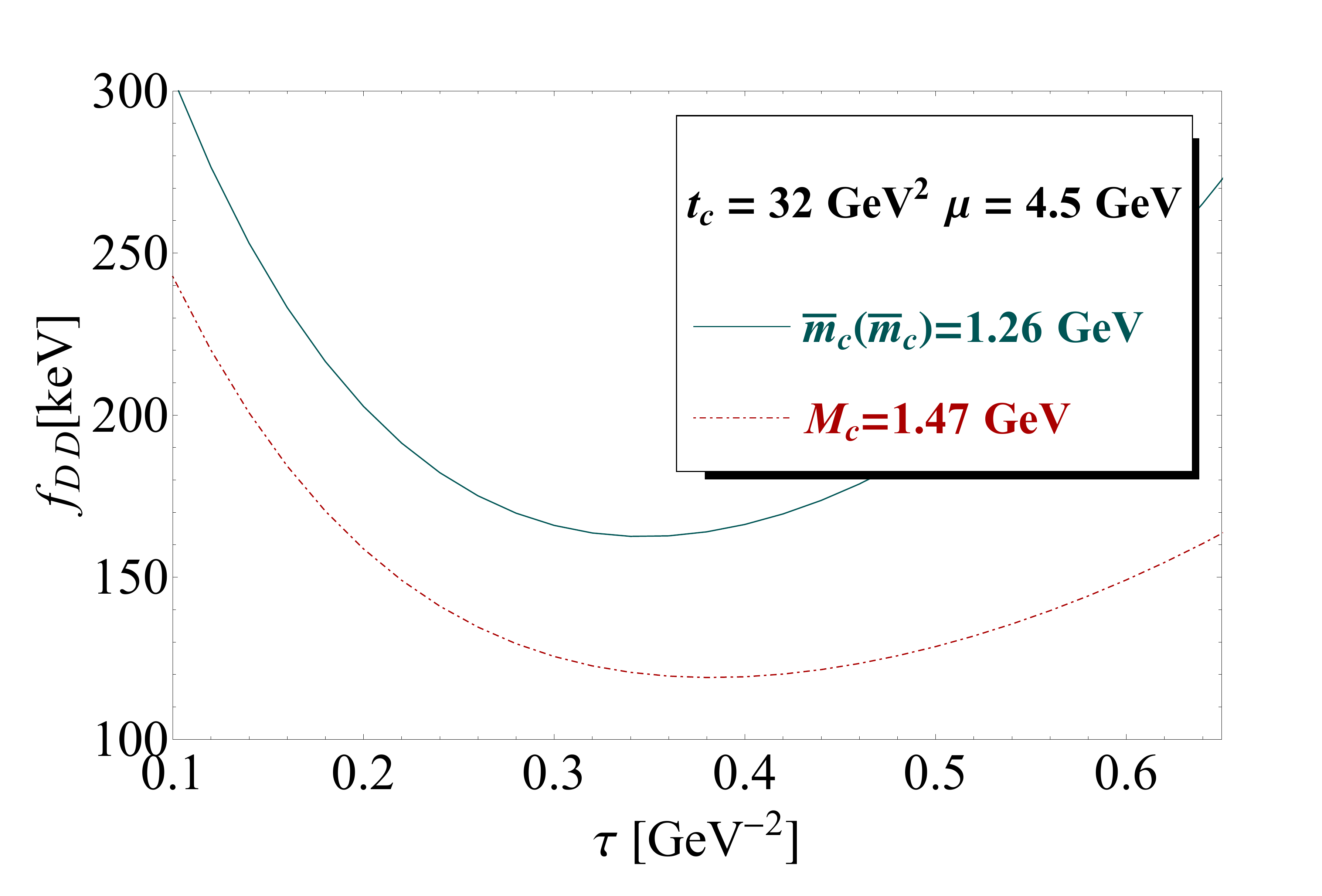}}
{\includegraphics[width=3.85cm  ]{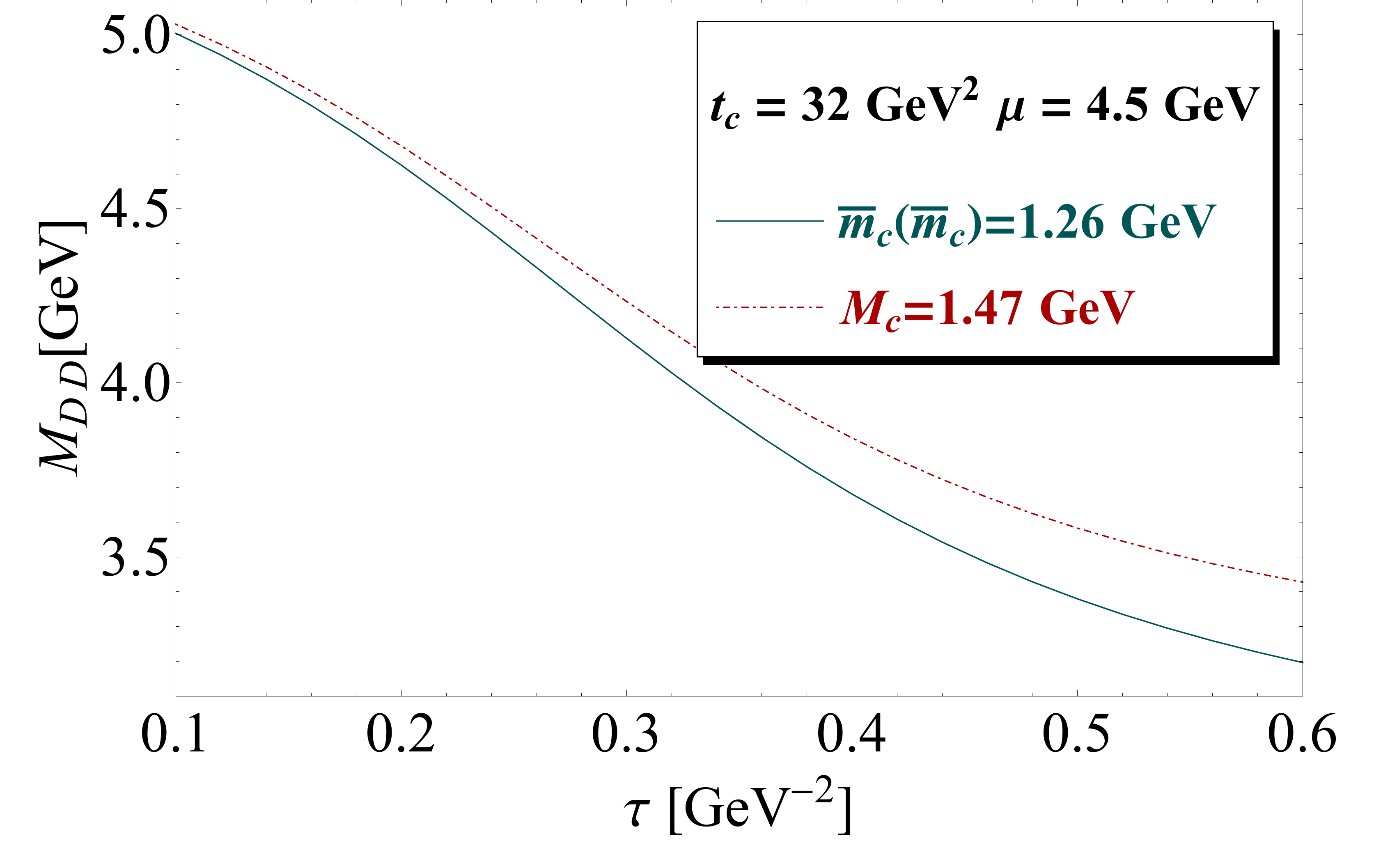}}
\scriptsize\centerline {\hspace*{-1cm} a)\hspace*{4cm} b) }
\caption{
\scriptsize 
{\bf a)} $f_{DD}$  at LO as function of $\tau$ for  $t_c=32$ GeV$^2$, for $\mu=4.5$ GeV, for  values of the running $\overline{m}_c(\overline{m}_c)=1.26$ GeV and pole mass $M_c=1.47$ GeV. We use     
the QCD parameters in Table\,\ref{tab:param}; {\bf b)} The same as a) but for the mass $M_{DD}$.
}
\label{fig:dd-const} 
\end{center}
\end{figure}

   \nin
   We show in Fig.\,\ref{fig:dd-const} the effect of the definitions (running and pole) of the heavy quark mass used in the analysis at LO which is relatively important. The difference should be added as errors in the LO analysis.  This source of errors is never considered in the current literature.
\nin
\subsubsection*{$\bullet$ Convergence of the PT series} 
\begin{figure}[hbt] 
\begin{center}
{\includegraphics[width=3.85cm  ]{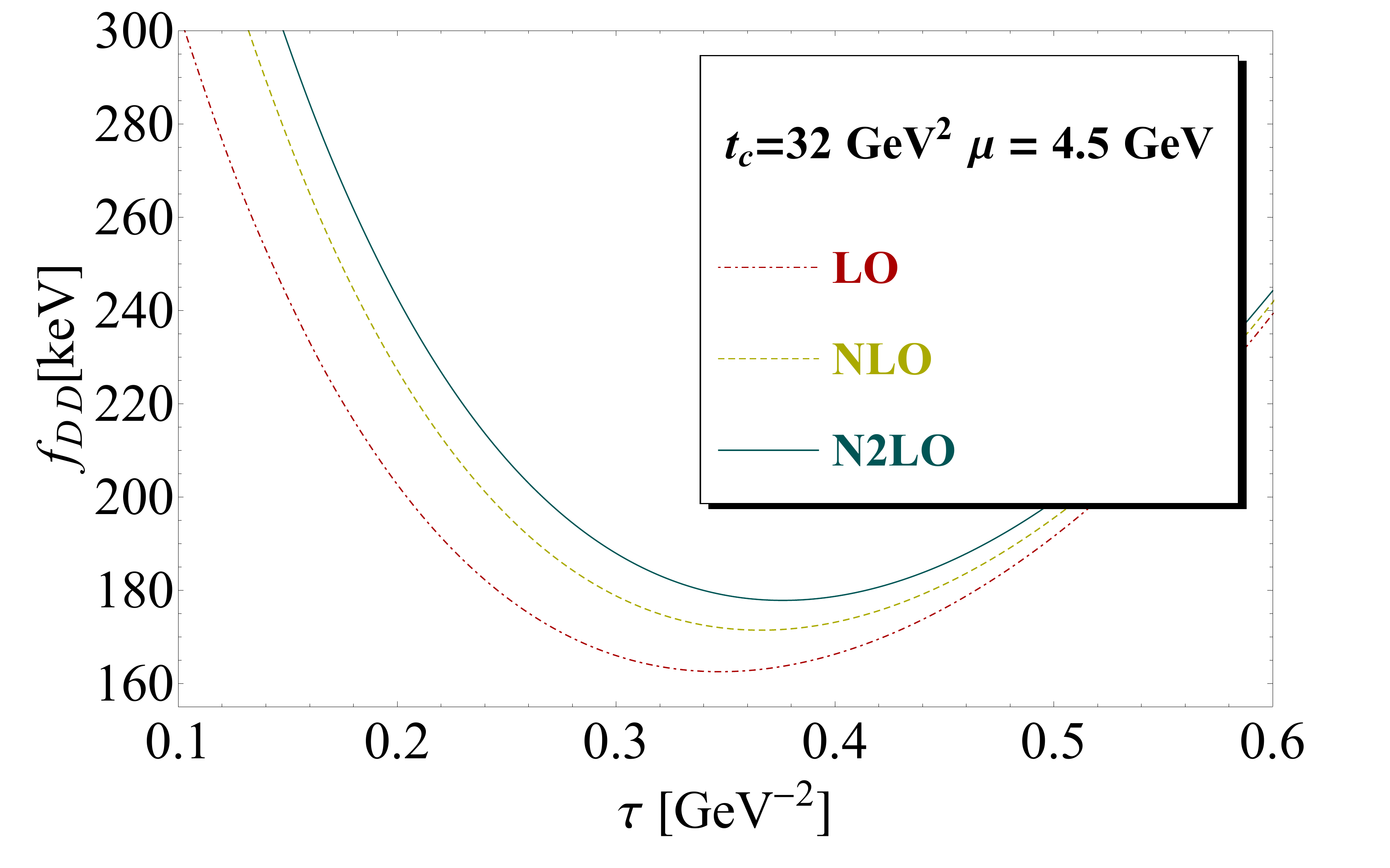}}
{\includegraphics[width=3.85cm  ]{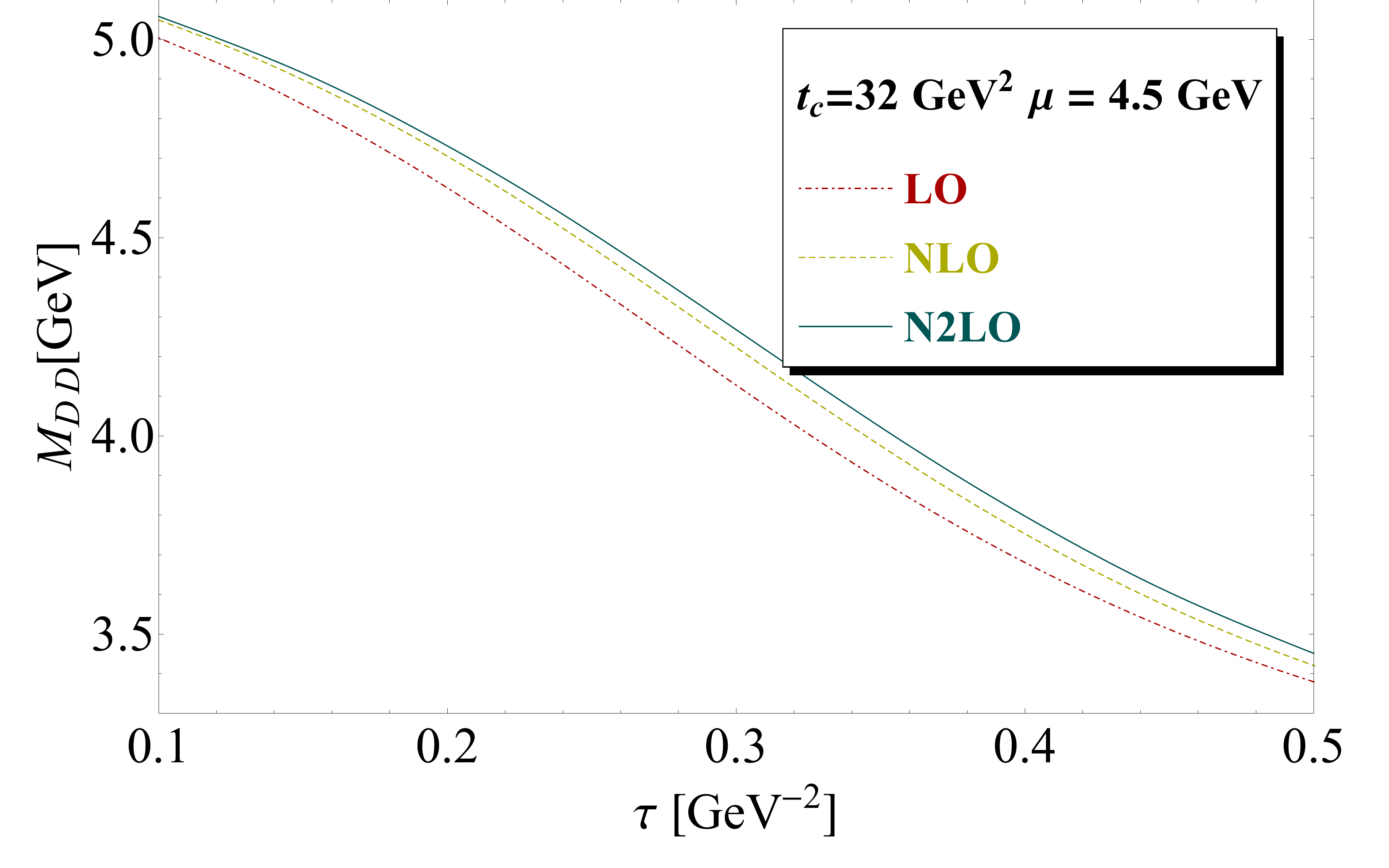}}
\scriptsize\centerline {\hspace*{-1cm} a)\hspace*{4cm} b)  }
\caption{
\scriptsize 
{\bf a)} $f_{DD}$  as function of $\tau$ for a given value of $t_c=32$ GeV$^2$, for $\mu=4.5$ GeV, for different truncation of the PT series  and for the QCD parameters in Table\,\ref{tab:param}; {\bf b)} The same as a) but for the mass $M_{DD}$.
}
\label{fig:d-pt} 
\end{center}
\end{figure}

\nin
Using  $t_c=32$ GeV$^2$, we study in Fig. {\ref{fig:d-pt}} the convergence of the PT series for a given value of $\mu=4.5$ GeV.  We observe (see Table\,\ref{tab:resultc}) that from NLO to N2LO the mass decreases by about only 1 per mil indicating the good convergence of the PT series.
\nin
\begin{figure}[hbt] 
\begin{center}
{\includegraphics[width=3.85cm  ]{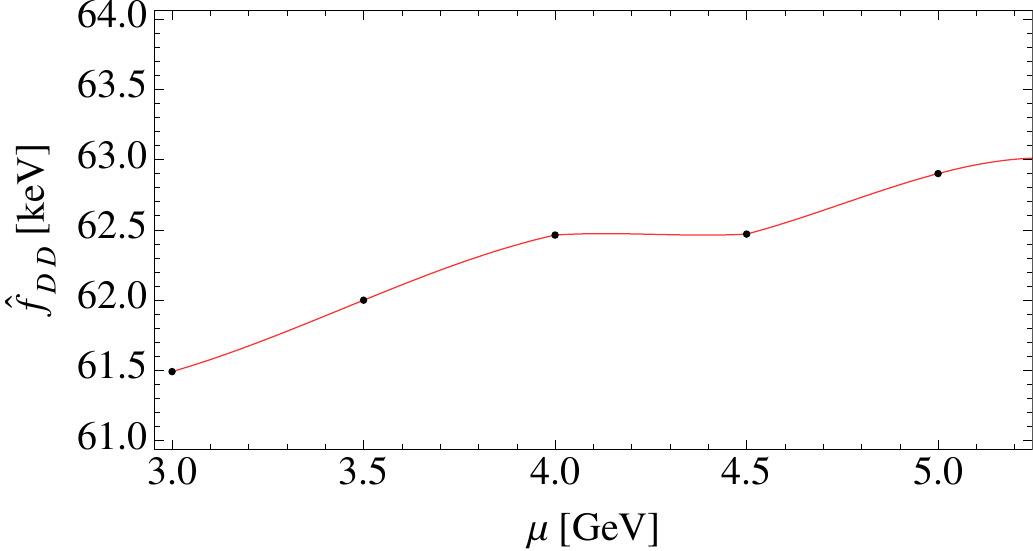}}
{\includegraphics[width=3.85cm  ]{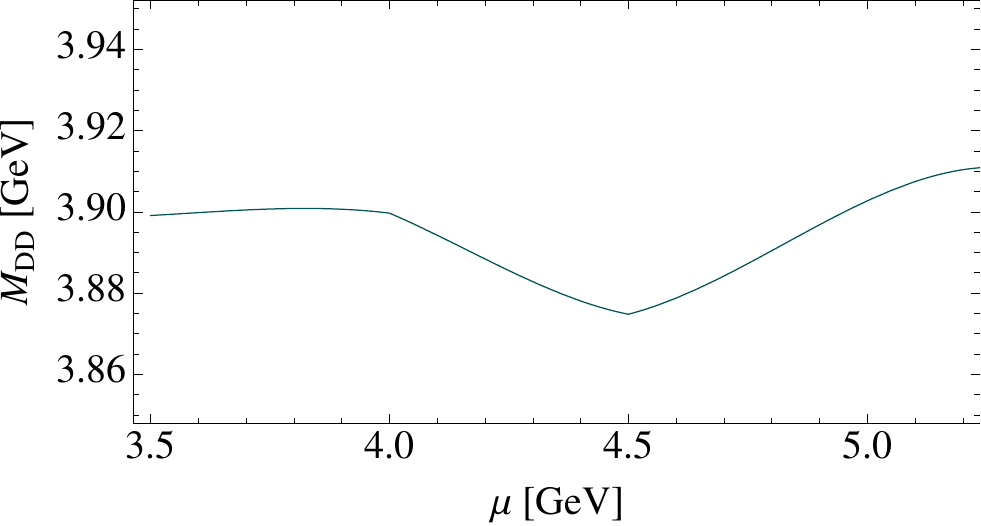}}
\scriptsize\centerline {\hspace*{-1cm} a)\hspace*{4cm} b)  }
\caption{
\scriptsize 
{\bf a)} Renormalization group invariant coupling $\hat f_{DD}$ at NLO as function of $\mu$, for the corresponding $\tau$-stability region, for $t_c\simeq 18$ GeV$^2$ and for the QCD parameters in Table\,\ref{tab:param};  {\bf b)} The same as a) but for the  mass $ M_{DD}$.
}
\label{fig:d-mu} 
\end{center}
\end{figure} 
\nin
\subsection*{$\bullet$ $\mu$-stability}
\nin
We improve our previous results by using different values of $\mu$ (Fig. {\ref{fig:d-mu}}). Using the fact that 
the final result must be independent of the arbitrary parameter $\mu$ (plateau / inflexion point for the coupling and minimum for the mass), we consider as an optimal result the one at  $\mu\simeq 4.5$ GeV where we deduce the result in Table\,\ref{tab:resultc}.
{\scriptsize
\begin{table*}[hbt]
\setlength{\tabcolsep}{0.8pc}
 \caption{$\bar DD$-like molecules masses, invariant and running couplings  from LSR within stability criteria at LO to N2LO of PT. 
 }  
\catcode`?=\active \def?{\kern\digitwidth}
{\scriptsize{
\begin{tabular*}{\textwidth}{@{}lll   lll  lll  l cc@{}}
\hline
\hline
                \bf Nature& \multicolumn{3}{c}{$\bf{\hat f_X}$ \bf [keV]} 
                 & \multicolumn{3}{c }{$\bf{ f_X(4.5)}$ \bf [keV]} 
                 &  \multicolumn{3}{c}{\bf Mass  [MeV]} 
                  &\bf Threshold
                 & \bf Exp.
                 \\
\cline{2-4} \cline{5-7}\cline{8-10}
                 & \multicolumn{1}{l}{LO} 
                 & \multicolumn{1}{l}{NLO} 
                 & \multicolumn{1}{l }{\bf N2LO} 
                      & \multicolumn{1}{l}{LO} 
                 & \multicolumn{1}{l}{NLO} 
                 & \multicolumn{1}{l }{\bf N2LO} 
                   & \multicolumn{1}{l}{LO} 
                 & \multicolumn{1}{l}{NLO} 
                 & \multicolumn{1}{l}{\bf N2LO} 
                  \\
\hline
 $\bf {J^{PC}=0^{++}}$ &&&&&&&&&&&--\\
$\bar DD$&56&60&\bf 62(6)&155&164&\bf 170(15)&3901&3901&\bf 3898(36)&3739
\\
$\bar D^*D^*$&--&--&--&269&288&\bf 302(47)&3901&3903&\bf 3903(179)&4020&\\
$ D^*_0D^*_0$&27&42&\bf 50(8)&74&116&\bf 136(22)&4405&4402&\bf 4398(54)&4636\\
\\
$\bf {J^{PC}=1^{+\pm}} $ &&&&&&&&&&&$X_c,Z_c$\\
$\bar D^*D$&87&93&\bf 97(10)&146&154&\bf 161(17)&3901&3901&\bf 3903(62)&3880\\
$\bar D^*_0D_1$&48&71&\bf 83(10)&81&118&\bf 137(16)&4394&4395&\bf 4401(164)&4739\\
\\
$\bf  {J^{PC}=0^{-\pm}} $ &&&&&&&&&&&--\\
$\bar D^*_0D$&68&88&\bf 94(7)&190&240&\bf 257(19)&5956&5800&\bf 5690(140)&4188\\
$\bar D^*D_1$&--&--&--&382&490&\bf 564( 38)&6039&5898&\bf 5797(141)&4432&\\
\\
$\bf { J^{PC}=1^{--}} $ &&&&&&&&&&&$Y_c$\\
$\bar D^*_0D^*$&112&143&$\bf 157(10)$&186&238&\bf 261(17) &6020&5861&\bf 5748(101)&4328\\
$\bar DD_1$&98&126&\bf 139(13)&164&209&\bf 231(21)&5769&5639&\bf 5544(162)&4291\\
$\bf { J^{PC}=1^{-+}} $ &&&&&&&&&&&$Y_c$\\
$\bar D^*_0D^*$&105&135&\bf 150(13)& 174 & 224 &\bf  249(22) &6047 & 5920 & \bf 5828(132)&4328 \\
$\bar DD_1$&97&128&\bf 145(15)&162&213&\bf 241(25)&5973&5840&\bf 5748 (179) \\
\\
\hline
\hline
\end{tabular*}
}}
\label{tab:resultc}
\end{table*}
}
\section{Molecule states masses and couplings}
The results are given in Table\,\ref{tab:resultc} (resp. Table\,\ref{tab:resultb}) for the charm (resp. bottom) channel where the corresponding hadronic threshold and experimental candidates are shown in the last two columns. The errors come from the QCD parameters and from the range of $\tau$, $t_c$ and $\mu$  where the optimal results are extracted.
{\scriptsize
\begin{table*}[hbt]
\setlength{\tabcolsep}{0.8pc}
 \caption{$\bar BB$-like molecules masses, invariant and running couplings  from LSR within stability criteria from LO to N2LO of PT. 
 }
\catcode`?=\active \def?{\kern\digitwidth}
{\scriptsize{
\begin{tabular*}{\textwidth}{@{}lll   lll  lll  l cc@{}}
\hline
\hline
                 \bf Nature& \multicolumn{3}{c}{$\bf{\hat f_X}$ \bf [keV]} 
                 & \multicolumn{3}{c }{$\bf{ f_X(5.5)}$ \bf [keV]} 
                 &  \multicolumn{3}{c}{\bf Mass  [MeV]} 
                  &\bf Threshold
                 & \bf Exp.
                 \\
\cline{2-4} \cline{5-7}\cline{8-10}
                 & \multicolumn{1}{l}{LO} 
                 & \multicolumn{1}{l}{NLO} 
                 & \multicolumn{1}{l }{\bf N2LO} 
                      & \multicolumn{1}{l}{LO} 
                 & \multicolumn{1}{l}{NLO} 
                 & \multicolumn{1}{l }{\bf N2LO} 
                   & \multicolumn{1}{l}{LO} 
                 & \multicolumn{1}{l}{NLO} 
                 & \multicolumn{1}{l}{\bf N2LO} 
                  \\
\hline
 $\bf{ J^{PC}=0^{++}}$ &&&&&&&&&&&--\\
$\bar BB$&4.0&4.4&\bf 5(1)&14.4&15.6&\bf 17(4)&10605&10598&\bf 10595(58)&10559\\
$\bar B^*B^*$&--&--&--&27&30&\bf 32(5)&10626&10646&\bf 10647(184)&10650\\
$ B^*_0B^*_0$&2.1&3.2&\bf 4(1)&7.7&11.3&\bf 14(4)&10653&10649&\bf 10648(113)&--\\
\\
$\bf { J^{PC}=1^{+\pm} }$ &&&&&&&&&&&$X_b,Z_b$\\
$\bar B^*B$&7&8&\bf 9(3)&14&16&\bf 17(5)&10680&10673&\bf 10646(150)&10605\\
$\bar B^*_0B_1$&4&6&\bf 7(1)&8&11&\bf 14(2)&10670&10679&\bf 10692(132)&--\\
\\
$\bf { J^{PC}=0^{-\pm}} $ &&&&&&&&&&&--\\
$\bar B^*_0B$&11&16&\bf 20(3)&39&55&\bf 67(10)&12930&12737&\bf 12562(260) &--
\\
$\bar B^*B_1$&--&--&-- &71&105&\bf 136(19) &12967&12794&\bf 12627(225)&11046\\
\\
$\bf  {J^{PC}=1^{--}} $ &&&&&&&&&&& $Y_b$\\
$\bar B^*_0B^*$&21&29&\bf 35(6) &39&54&\bf 66(11) &12936&12756&\bf 12592(266) &--\\
$\bar BB_1$&21&29&\bf 35(7)&39&54&\bf 65(12)&12913&12734&\bf 12573(257)&11000\\
$\bf  {J^{PC}=1^{-+}} $ &&&&&&&&&&& $Y_b$\\
$\bar B^*_0B^*$&20&29&\bf 34(4)&38&54&\bf 64(8)&12942&12774&\bf 12617(220) &-- \\
$\bar BB_1$& 20&29&\bf 35(5)&	37&53&\bf 65(9) &12974& 12790&\bf 12630(236)					&11000\\

\hline
\hline
\end{tabular*}
}}
\label{tab:resultb}
\end{table*}
}
{\scriptsize
\begin{table*}[hbt]
\setlength{\tabcolsep}{1.pc}
\caption{Four-quark masses, invariant and running couplings  from LSR within stability criteria from LO to N2LO of PT. 
}  
\catcode`?=\active \def?{\kern\digitwidth}
{\scriptsize{
\begin{tabular*}{\textwidth}{@{}lll   lll  lll  l c@{}}
\hline
\hline
                \bf Nature& \multicolumn{3}{c}{\bf{$\hat f_X$} \bf [keV]} 
                 & \multicolumn{3}{c }{\bf{ $f_X(4.5)$} \bf [keV]} 
                 &  \multicolumn{3}{c}{\bf Mass  [MeV]} 
                 & Exp.
                 \\
\cline{2-4} \cline{5-7}\cline{8-10}
                 & \multicolumn{1}{l}{LO} 
                 & \multicolumn{1}{l}{NLO} 
                 & \multicolumn{1}{l }{\bf N2LO} 
                      & \multicolumn{1}{l}{LO} 
                 & \multicolumn{1}{l}{NLO} 
                 & \multicolumn{1}{l }{\bf N2LO} 
                   & \multicolumn{1}{l}{LO} 
                 & \multicolumn{1}{l}{NLO} 
                 & \multicolumn{1}{l}{\bf N2LO} 
                  \\
\hline
 \bf{$c$}\bf-quark \\
  $S_c(0^{+})$&62&67&\bf 70(7)&173&184&\bf 191(20)&3902&3901&\bf 3898(54)&--  \\
    $A_c(1^{+})$&100&106&\bf 112(18)&166&176&\bf 184(30)&3903&3890&\bf 3888(130)&$X_c,Z_c$\\
    $\pi_c(0^{-})$&84&106&\bf 113(5)&233&292&\bf 310(13)&6048&5872&\bf 5750(127)&--  \\
         $V_c(1^{-})$&123&162&\bf 178(11)&205&268&\bf 296(19)&6062&5904&\bf 5793(122)&$Y_c$\\
 \hline
\hline
\end{tabular*}
}}
\label{tab:4q-resultc}
\end{table*}
}

{\scriptsize
\begin{table*}[hbt]
\setlength{\tabcolsep}{1.pc}
\caption{Four-quark masses, invariant and running couplings from LSR within stability criteria from LO to N2LO of PT. 
}  
\catcode`?=\active \def?{\kern\digitwidth}
{\scriptsize{
\begin{tabular*}{\textwidth}{@{}lll   lll  lll  l c@{}}
\hline
\hline
                \bf Nature& \multicolumn{3}{c}{\bf{$\hat f_X$} \bf [keV]} 
                 & \multicolumn{3}{c }{\bf{ $f_X(5.5)$} \bf [keV]}  
                 &  \multicolumn{3}{c}{\bf Mass  [MeV]}                  
                 & Exp.
                 \\
\cline{2-4} \cline{5-7}\cline{8-10}
                 & \multicolumn{1}{l}{LO} 
                 & \multicolumn{1}{l}{NLO} 
                 & \multicolumn{1}{l }{\bf N2LO} 
                      & \multicolumn{1}{l}{LO} 
                 & \multicolumn{1}{l}{NLO} 
                 & \multicolumn{1}{l }{\bf N2LO} 
                   & \multicolumn{1}{l}{LO} 
                 & \multicolumn{1}{l}{NLO} 
                 & \multicolumn{1}{l}{\bf N2LO} 
                  \\
\hline
\bf {$b$}\bf-quark \\
   $S_b(0^{+})$&4.6&5.0&\bf 5.3(1.1)&16&17&\bf 19(4)&10652&10653&\bf 10654(109)&--  \\
    $A_b(1^{+})$&8.7&9.5&\bf 10(2)&16&18&\bf 19(3)&10730&10701&\bf 10680(172)&$Z_b$\\ 
 $\pi_b(0^{-})$&18&23&\bf 27(3)&62&83&\bf 94(11)&13186&12920&\bf 12695(254)&--  \\
 $V_b(1^{-})$&24&33&\bf 40(5)&45&62&\bf 75(9)&12951&12770&\bf 12610(242)&$Y_b$\\

\hline
\hline
\end{tabular*}
}}
\label{tab:4q-resultb}
\end{table*}
}
\section{Four-quark states masses and couplings}
The results are given in Table\,\ref{tab:4q-resultc} (resp. Table\,\ref{tab:4q-resultb}) for the charm (resp. bottom) channel where the experimental candidates are shown in the last column. The sources of errors are the same as in the molecules case.
\section{Confrontation with data and some LO results}\label{sec:confront}
\subsection*{\b Axial-vector $(1^{++})$ states}
As mentioned in the introduction, there are several observed states in this channel.
 In addition to the well-established $X_c(3872)$, we have the $X_c(4147,4273)$ and the $Z_c(3900,4025,4050,4430)$.

 For the non-strange states found from their decays into $J/\psi \pi^+\pi^-$, one can conclude from the results given in Table\,\ref{tab:resultc}  that the $X_c(3872)$ and $Z_c(3900)$
can be well described with an almost pure  $\bar D^*D$ molecule or/and four quark $[cq\bar c\bar q]$  states,  ($q\equiv u,d$) while the one of the $Z_c(4200,4430)$ might be a $\bar D^*_0D_1$ molecule state. Our results for the $X_c(3872)$ confirm our LO ones in\,\cite{X1A,X1B,X2}. 

One can notice that the values of these masses below the corresponding $\bar DD,\bar BB$-like thresholds are much lower than the ones predicted $\simeq 5.12$ (resp $11.32$) GeV for the $1^{++}$ $\bar cgc$ (resp. $\bar bgb$) hybrid mesons\,\cite{KLEIV1b,GOVAERTS,GOVAERTS2,SNB2}. 

 Assuming that the value of $\sqrt{t_c}\approx (6-7)$ GeV, where the optimal values of the masses have been extracted, are approximately the mass of the 1st radial excitation, one can deduce that the higher masses experimental states cannot be such radial excitations. 

 In the bottom sector, experimental checks of our predictions are required. 

\subsection*{\b Scalar $(0^{++})$ states}
Our analysis in Tables\,\ref{tab:resultc} and \ref{tab:4q-resultb} predicts that:

 The $0^{++}$ $\bar DD,\bar D^*D^*$ molecule  and four-quark non-strange states are almost degenerated with the $1^{++}$ ones and have masses around 3900 MeV. This prediction is comparable with the $Z_c(3900)$ quoted by PDG\,\cite{PDG} as a $0^{++}$ state. 

 The predicted mass of the $\bar D^*_{0}D^*_{0}$ molecule is higher [4402(30) MeV] but is still below the $\bar D^*_{0}D^*_{0}$ threshold. 

\subsection*{\b Vector $(1^{-\pm})$ states}
Our predictions in Tables\,\ref{tab:resultc} to \ref{tab:4q-resultb} for molecules  and four-quark vector states in the range of (5646-5961) MeV are too high compared with the observed $Y_c(4140)$ to $Y_c(4660)$ states. Our N2LO results confirm previous LO ones in\,\cite{X3A,RAFMOLE1} but
do not support the result in \,\cite{MOLE12b} which are too low. 

Our results  indicate that the observed  states might result from a mixing of the molecule / four-quark  with ordinary quarkonia-states (if the description of these states in terms of molecules and/or four-quark states are the correct one). The NP contribution to this  kind of mixing has been estimated to leading order in\,\cite{RAPHAEL1}. The same conclusion holds for the $Y_b(9898,10260,10870)$ where the predicted unmixed molecule / four-quark states are in the range (12326-12829) MeV. 

As these pure molecule states are well above the physical threshold, they might not be bound states and could not be separated from backgrounds.  Our  results go in lines with the ones of\,\cite{ZAHED}. 
\subsection*{\b Pseudoscalar $(0^{-\pm})$ states}
\nin
One expects from Tables\,\ref{tab:resultc} to \ref{tab:4q-resultb} that the $0^{-\pm}$ molecules will populate the region 5656-6020 (resp 12379-12827) MeV for the charm (resp bottom) channels like in the case of the $1^{-\pm}$ vector states. One can notice that these states are much heavier than the predicted
 $0^-$ hybrid $\bar cgc$ (resp. $\bar bgb$) ones 
$\simeq  3.82$ (resp. $\simeq 10.64$) GeV from QSSR\,\cite{KLEIV2b,GOVAERTS,GOVAERTS2,SNB2}.  Like in the case of vector states, these pseudoscalar states are well above the physical threshold. Therefore, like in the case of vector states, these molecule states should be broad and are difficult to separate from backgrounds. 

One can also notice that the $D^*_0D(0^{--})$ and $(0^{-+})$ states are almost degenerate despite the opposite signs of the $\qq$ and $\la \bar qGq\ra$ contributions to  the spectral functions in the two channels (see Appendix of \cite{XYZ}).
\subsection*{\b Isospin breakings and almost degenerate states}
In our approach, isospin breakings are controlled by the running light quark mass  $\bar m_d-\bar m_u$ and condensate $\la\bar uu-\bar dd\ra$ differences which are tiny quantities. Their effects are hardly noticeable within the accuracy of our approach. Therefore, for the neutral combination of currents which we have taken in Table\,\ref{tab:current}, one expects that the molecules built from the corresponding charged currents will be degenerate in masses because their QCD expressions are the same in the chiral limit. 
\subsection*{\b Radial excitations}
If one considers the value of the continuum threshold $t_c$, at which the optimal value of the ground state is obtained, 
as an approximate value of the mass of the 1st radial excitation, one expects that the radial excitations are in the region of about 0.4 to 1.6 GeV above the ground state mass. A more accurate prediction can be obtained by combining LSR with Finite Energy Sum Rule (FESR)\,\cite{X1A,X1B,X3A} where the mass-splitting is expected to be around 250-300 MeV at LO. Among these different observed states, the $Z_c(4430)$ and $X_c(4506,4704)$ could eventually be considered as radial excitation candidates. 
\section{Quark Mass Behaviour of the Decay Constants }\label{sec:coupling}
The couplings or decay constants given in Tables\,\ref{tab:resultc} to \ref{tab:4q-resultb} are normalized
in Eq.\,\ref{eq:coupling} in the same way as $f_\pi=130.4(2)$ MeV through its coupling to the pseudoscalar 
current :  $\la 0|(m_u+m_d)\bar u(i\gamma_5)d|\pi\ra=f_\pi m_\pi^2 \phi_\pi(x)$, where $\phi_\pi(x)$  is the pion field.

One can find from Table\,\ref{tab:resultc} that $f_{DD}\simeq 170(15)$ keV which is about $10^{-3}$ of 
$f_\pi$ and of $f_B\simeq f_D\simeq$ 206(7) MeV\,\cite{SNFB12a,SNREV15,SNFB14}. The same observation holds for the other molecule and four-quark states indicating the weak coupling of these states to the associated interpolating currents. 

 Comparing the size of the couplings in the $c$ and $b$ quark channels (Tables\,\ref{tab:resultc} to \ref{tab:4q-resultb}), one can observe that the ratio decreases by a factor about 10 from the $c$  to the $b$ channels for the $0^{++}$ and $1^{++}$ states  which is about the value of the ratio $(\bar m_c/\bar m_b)^{3/2}$, while it decreases by about a factor 4  for the $0^{--}$ and $1^{--}$ states  which is about the value $(\bar m_c/\bar m_b)$.  These behaviours can be compared with the well-known one of $f_B\sim 1/\bar m_b^{1/2}$ from HQET and can motivate further theoretical studies of the molecule and four-quark couplings. 
 {\scriptsize
\begin{table}[hbt]
 \caption{Exotic hadron masses and couplings from LSR within stability at N2LO}  
\setlength{\tabcolsep}{0.25pc}
    {\footnotesize
 {\begin{tabular}{@{}lllll@{}} 
&\\
\hline
\hline
\bf Nature&\bf$J^{P}$&\bf Mass [MeV] & \bf $\hat f_X$ [keV] & \bf $ f_X(4.5)$ [keV]  \\
\hline
{\bf  $b$-quark channel}&\\
{\it Molecule} &&&\\
$B^*K$&$1^{+}$&$5186\pm 13$ &$4.48\pm 1.45$&$8.02\pm 2.60$ \\
$BK$&$0^{+}$&$5195\pm 15$&$2.57\pm 0.75$&$8.26\pm 2.40$\\
$ B^*_s\pi$&$1^{+}$&$5200\pm 18$&$5.61\pm 0.87$&$10.23\pm 1.59$\\
$ B_s\pi$&$0^{+}$&$5199\pm 24$&$3.15\pm 0.70 $&$10.5\pm 2.30$\\
{\it Four-quark $(su)(\bar b\bar d)$} &&&\\
$A_b$&$1^{+}$&$5186\pm 16$&$5.05\pm 1.32$&$9.04\pm 2.37$\\
$S_b$&$0^{+}$&$5196\pm 17$&$2.98\pm 0.70$&$9.99\pm 2.36$\\
{\bf  $c$-quark channel}&\\
{\it Molecule}\\
$ D^*K$&$1^{+}$&$2395\pm 48$ &$155\pm 36$&$226\pm 52$ \\
$ DK$&$0^{+}$&$2402\pm 42$&$139\pm 26$&$254\pm 48$\\
$ D^*_s\pi$&$1^{+}$&$2395\pm 48$ &$215\pm 35$&$308\pm 49$ \\
$ D_s\pi$&$0^{+}$&$2404 \pm 37$&$160 \pm 22$&$331 \pm 46$\\

{\it Four-quark $(su)( \bar c\bar d)$} &&&\\
$A_c$&$1^{+}$&$2400\pm 47$&$192\pm 41$&$260\pm 55$\\
$S_c$&$0^{+}$&$2395\pm 68$&$122\pm 26$ &$221\pm 47$\\
\hline\hline
\end{tabular}}
\label{tab:resultx}
}
\end{table}
} 
\section{The case of the $X(5568)$}
 In\,\cite{X}, we have also studied the $X$ hadron formed by 3 light quarks $uds$ and one heavy quark $Q\equiv c,b$ using the same approach as above by assuming if it is a  molecule or four-quark state.  We have included NLO andN2LO PT corrections and the contributions of condensates of dimension $d\leq 7$.  Our results are summarized in Table\,\ref{tab:resultx}. Contrary to previous claims in the sum rule literature, our results do not favour a $BK,~BK^*$ or $B_s\pi$ molecule or four-quark $(bu)( \bar d \bar s)$ state having a mass around 5568 MeV observed by D0\,\cite{D0X} but not confirmed by LHCb\,\cite{LHCbX}. We also predict the corresponding state in the $c$-quark channel where the $D^*_{s0}(2317)$ seen by BABAR\,\cite{BABAR} in the $D_s\pi$ invariant mass, expected to be an isoscalar-scalar state with a width less than 3.8 MeV \,\cite{PDG}  could be a good candidate for one of such states.
\section{Conclusions}
We have presented in these talks a summary of the results obtained in the chiral limit at N2LO of PT\,\cite{XYZ,X}. The extension of this work including $SU(3)$ breaking terms which we shall compare with recent experimental states decaying to  $J/\psi\phi$ is under investigation.


\begin{thebibliography}{999}
\bibitem{EXP} More details and complete references are in Section 1 of \cite{XYZ}.
\bibitem{TH} Some related theoretical works are quoted in\,\cite{XYZ} and\,\cite{X}.
\end{thebibliography}
\end{document}